\documentclass[12pt,a4paper,twoside]{article}

\usepackage{authblk}

\usepackage[utf8]{inputenc}   

\usepackage[english]{babel} 

\usepackage[T1]{fontenc}

\usepackage{geometry}	
\usepackage{setspace}

\usepackage{graphicx}

\usepackage{amsmath}  
\usepackage{amsthm}   
\usepackage{amsfonts} %
\usepackage{amssymb}

 \usepackage{wrapfig}

\usepackage{subfigure}

\usepackage{subfigmat}

 \usepackage{url}

\usepackage{varioref}

\usepackage{dcolumn}
\newcolumntype{d}{D{.}{.}{-1}} 
\newcolumntype{e}{D{E}{E}{-1}} 

 \usepackage{verbatim}

\usepackage{rotating}

\usepackage[pdftex]{hyperref} 
\hypersetup{colorlinks,       
            linkcolor=black,  
            anchorcolor=black,
            citecolor=black,  
            filecolor=black,  
            menucolor=black,  
            pagecolor=black,  
            urlcolor=black,   
	          bookmarks=true,         
	          bookmarksopen=false,    
	          bookmarksnumbered=true, 
	         }

\usepackage[figure,table]{hypcap}
\usepackage{float}

\usepackage[numbers,sort&compress]{natbib} 

\usepackage{notoccite}

\usepackage{multirow}

\usepackage{booktabs}

\usepackage{pdfpages}

\usepackage{csquotes}
\usepackage{nameref}  

\usepackage{tikz}
\usetikzlibrary{positioning}

\usepackage{ragged2e}  

\usepackage{enumitem}

\usepackage{diagbox}

\usepackage{appendix}

\usepackage{nicefrac, xfrac}

\usepackage{makecell}  

\usepackage{colortbl}

\usepackage[bf,small]{caption}

\DeclareCaptionType{equ}[][]
\usepackage[export]{adjustbox}

\usepackage{hhline}

\usepackage{fancyhdr}

\usepackage{array}

\usepackage{graphicx}

\title{Collective dynamics of strategic classification}

\author[1]{Marta C. Couto\thanks{m.gomesdacunhacouto@uva.nl}}
\author[2,3]{Flavia Barsotti\thanks{flavia.barsotti@ing.com, f.barsotti@tudelft.nl | Corresponding author. ING Research Project Lead. For the authors Flavia Barsotti and Fernando P. Santos names are reported in alphabetical order.}} 
\author[1]{Fernando P. Santos\thanks{f.p.santos@uva.nl | Corresponding author}}

\affil[1]{Informatics Institute, University of Amsterdam, The Netherlands}
\affil[2]{ING Analytics, ING Group, Amsterdam, The Netherlands}
\affil[3]{Delft Institute of Applied Mathematics (DIAM), TU Delft, The Netherlands}

\date{\today}

\begin{document}

\maketitle

\begin{abstract}	

 Classification algorithms based on Artificial Intelligence (AI) are nowadays applied in high-stakes decisions in finance, healthcare, criminal justice, or education.  
Individuals can strategically adapt to the information gathered about classifiers, which in turn may require algorithms to be re-trained.  Which collective dynamics will result from users' adaptation and algorithms' retraining? We apply evolutionary game theory to address this question. Our framework provides a mathematically rigorous way of treating the problem of feedback loops between collectives of users and institutions, allowing to test interventions to mitigate the adverse effects of strategic adaptation. 
As a case study, we consider institutions deploying algorithms for credit lending. We consider several scenarios, each representing different interaction paradigms. When algorithms are not robust against strategic manipulation, we are able to capture previous challenges discussed in the strategic classification literature, whereby users either pay excessive costs to meet the institutions' expectations (leading to high social costs) or game the algorithm (e.g., provide fake information). From this baseline setting, we test the role of improving gaming detection and providing algorithmic recourse. We show that increased detection capabilities reduce social costs and could lead to users' improvement; when perfect classifiers are not feasible (likely to occur in practice), algorithmic recourse can steer the dynamics towards high users' improvement rates.
The speed at which the institutions re-adapt to the user's population plays a role in the final outcome. Finally, we explore a scenario where strict institutions provide actionable recourse to their unsuccessful users and observe cycling dynamics so far unnoticed in the literature.

\vspace{3mm}
\end{abstract}

\section*{Significance Statement}

Artificial Intelligence (AI) is increasingly supporting high-stakes decisions such as credit approvals, hiring, and fraud detection. 
In these contexts, users may decide to adapt their behaviour or application materials, based on information about how algorithms operate. Adaptation can involve meaningful improvement, but can also include strategic adaptation and gaming (e.g., faking information - misreporting qualifications or financial data). 
 
These interactions create a feedback loop between users and institutions that can compromise performance, accuracy, and fairness of algorithms over time.
This study presents a novel theoretical framework capturing the long-run dynamics of these interactions by means of evolutionary game theory. 
In contrast to earlier works, which typically model users as fully informed and decisions as one-off events, our framework accounts 
for repeated adaptation, partial knowledge, and the presence of multiple institutions with different acceptance thresholds. This enables a more realistic understanding of the dynamics underlying the feedback loop between institutions and users. 
Our results highlight how imperfect classifiers can lead to long-term social costs, inducing individuals to over-adapt. We show that improving gaming detection or providing algorithmic recourse can alleviate the problems of strategic classification. Counterintuitively, we also highlight that faster algorithmic re-training can lead to undesirable states where gaming by unqualified users prevails.  

By modelling the mutual adaptation of users and institutions, this work contributes to a deeper understanding of collective dynamics in sociotechnical systems where algorithms are widespread, offering insights on designing AI systems that are not only accurate and robust but also socially responsible.

\section{Introduction}

Artificial Intelligence (AI) and Machine Learning-aided decisions are ubiquitous nowadays.
When they concern high-stakes domains, such as credit lending, insurance, fraud detection, hiring, or university admission, algorithmic decisions can severely impact individuals' well-being \cite{Raghavan:book:2023} and potentially be harmful. Regulation is stressing the importance of a responsible AI development and use \cite{EUAIA24}. 
As a result, algorithms should be increasingly transparent and actionable, providing both the information and the means for users to reach a positive classification. 
Therefore, institutions deploying the algorithms simultaneously aim to provide an effective service (e.g., accepting clients as loan borrowers that have a high chance of repayment and declining risky ones) and to be transparent (for ethical and legal motivations) \cite{Goodman:AIMag:2017, Wachter:HarvJLTech:2018}. 

When users can anticipate how their data will be used in algorithms --- which is more likely to happen as algorithms become increasingly transparent --- they may alter their actions or features to receive favourable classification outcomes. Adaptation is desirable when it leads to honest improvement. On the other hand, individuals might aim at \textit{gaming} the system and manipulating data (e.g., providing fake information on savings or skills) rather than trying to improve \cite{Akyol:arXiv:2016, Bambauer:NDLR:2018, Kearns:bookchapter:2020}. 
The challenge of designing algorithms to cope with strategic adaptation to influence a given classification outcome is called \textit{strategic classification} \cite{Hardt:ITCS:2016}.

Knowledge about the classification algorithms can introduce data distribution shifts, affecting the accuracy of model predictions \cite{Perdomo:ICML:2020, Tsvetkova:NatHumBehav:2024, Dean:TMLR:2025}.
To mitigate this risk and its impacts on the algorithmic performance and accuracy, institutions can impose an additional burden on applicants, in terms of requirements during the application process for example. 

Therefore, strategic behaviour creates a dynamic feedback loop between users and institutions.
Ignoring these complex dynamics can lead to lower model performance for institutions and eventually high social costs for users \cite{Milli:FAT:2019, Hu:FAT:2019, Bechavod:PMLR:2022}. 
It is essential for institutions to design models that not only classify accurately but also take into account the risks deriving from potential data distribution shifts. In some cases, these shifts can derive from evolving strategic responses of users. 
But how do individuals adapt in practice? 
And how does the co-evolving process of users' adaptation and algorithms' re-training impact both people and institutions deploying classifiers in the long run? 
Such questions are fundamental in banking and finance applications, as individuals can adapt their behaviour to game predictive models used for credit scoring, loan approval, or fraud detection.
However, these questions remain under-explored in the current literature.
Here, we argue that mathematical models are crucial in helping us understand the collective dynamics of the corresponding sociotechnical complex systems \cite{Chopra:AIES:2018, Murukannaiah:AAMAS:2020, Tsvetkova:NatHumBehav:2024, Dean:TMLR:2025}. 

In their seminal paper \cite{Hardt:ITCS:2016}, Hardt et al. introduce the problem of strategic classification. 
They focus on its impact on model performance, asking whether one can design strategy-robust classifiers.
Instead, other related work asks whether and how classifiers can incentivize honest improvement \cite{Kleinberg:TEAC:2020}.
Both studies frame the interaction between people and institutions as a game --
the institutions can decide which algorithms to deploy, and the users decide how to respond to those, e.g., which of their relevant features to disclose or alter.
Importantly, they assume that the users perfectly know the classifier, and can compute their best response to that.
In turn, institutions use this assumption on users to possibly define their strategy-robust algorithm or create incentives for improvement.

Later works aim at modelling user behaviour more realistically, by considering noisy learning \cite{Jagadeesan:PMLR:2021, Zrnic:ANeurIPS:2021, Haghtalab:EC:2022}, or social learning \cite{Bechavod:PMLR:2022,Barsotti:AAMAS:2022, Barsotti:IJCAI:2022}.
For example, Barsotti et al. (2022) assume a population of users where, besides individual optimization, they can decide to copy other individuals’ behavior via an imitation process \cite{Barsotti:AAMAS:2022,Barsotti:IJCAI:2022}. Imitation can potentially mitigate some negative impacts on classification. The more agents rely on imitation, the lower are the negative impacts on classification. 
The proposed setting in \cite{Barsotti:AAMAS:2022,Barsotti:IJCAI:2022} allows to study the effects of imitation for different levels of transparency in the system, referring to the information available to users about the classifiers. 
Additionally, some other works focus on the effect of various levels of algorithmic transparency \cite{Akyol:arXiv:2016, Ghalme:ICML:2021, 
Shao:arXiv:2025}.
Interestingly, while transparency may leave room for 
strategic behavior, full opaqueness can also be detrimental, even for institutions \cite{Ghalme:ICML:2021}.
Most of these previous studies share a common constraint: they model the interaction as a single-step process, in which an institution deploys an algorithm and users respond once.
This simplification overlooks the prolonged and dynamic nature of the relation between AI algorithms and their users 
--- one that typically unfolds over multiple rounds of continuous 
adaptation \cite{Dean:TMLR:2025}.
Here, we address this gap.

We develop a general theoretical framework that models the interaction between institutions and users over time, reflecting the idea of human-AI co-evolution \cite{pedreschi2025human}. 
For that, we abstract from the algorithmic details and take an evolutionary game theory approach \cite{Hofbauer:book:1998}.
Evolutionary game theory, although initially proposed in behavioural biology \cite{Smith:Nature:1973}, has proven to be a remarkably useful framework to model sociotechnical dynamics, such as AI race regulation \cite{Han:JAIR:2020}, technology adoption \cite{Encarnacao:TRPB:2018}, multi-sector coordination \cite{Encarnacao:RSO:2016}, or forecasting in dynamic environments \cite{Tilman:CollectInt:2023}.
To our knowledge, only one other work on strategic classification takes a similar approach to ours in this respect \cite{Saig:arXiv:2025}.
Saig and Rosenfeld (2025) study the long-term co-evolution of model performance and user population composition, specifically using the replicator dynamics. The authors consider the setting of a single institution (the learner) and a population of users evolving through \textit{natural selection}, assuming that fitness depends on how adequately the institution classifies users \cite{Saig:arXiv:2025}. In contrast, we consider both a population of users and a population of institutions. 
This accounts for direct competition among institutions, a feature that is absent in the previous work.
Additionally, Saig and Rosenfeld (2025) assume that the institution's and users' utilities are fully aligned.
In particular, the users always benefit from a good performance of the classification algorithm.
In alternative, we consider potential conflicts of interest between users and institutions, 
an aspect relevant in practice when unqualified users can benefit from low algorithmic precision.

\begin{figure}[p!]
    \centering
    \includegraphics[width=1\linewidth]{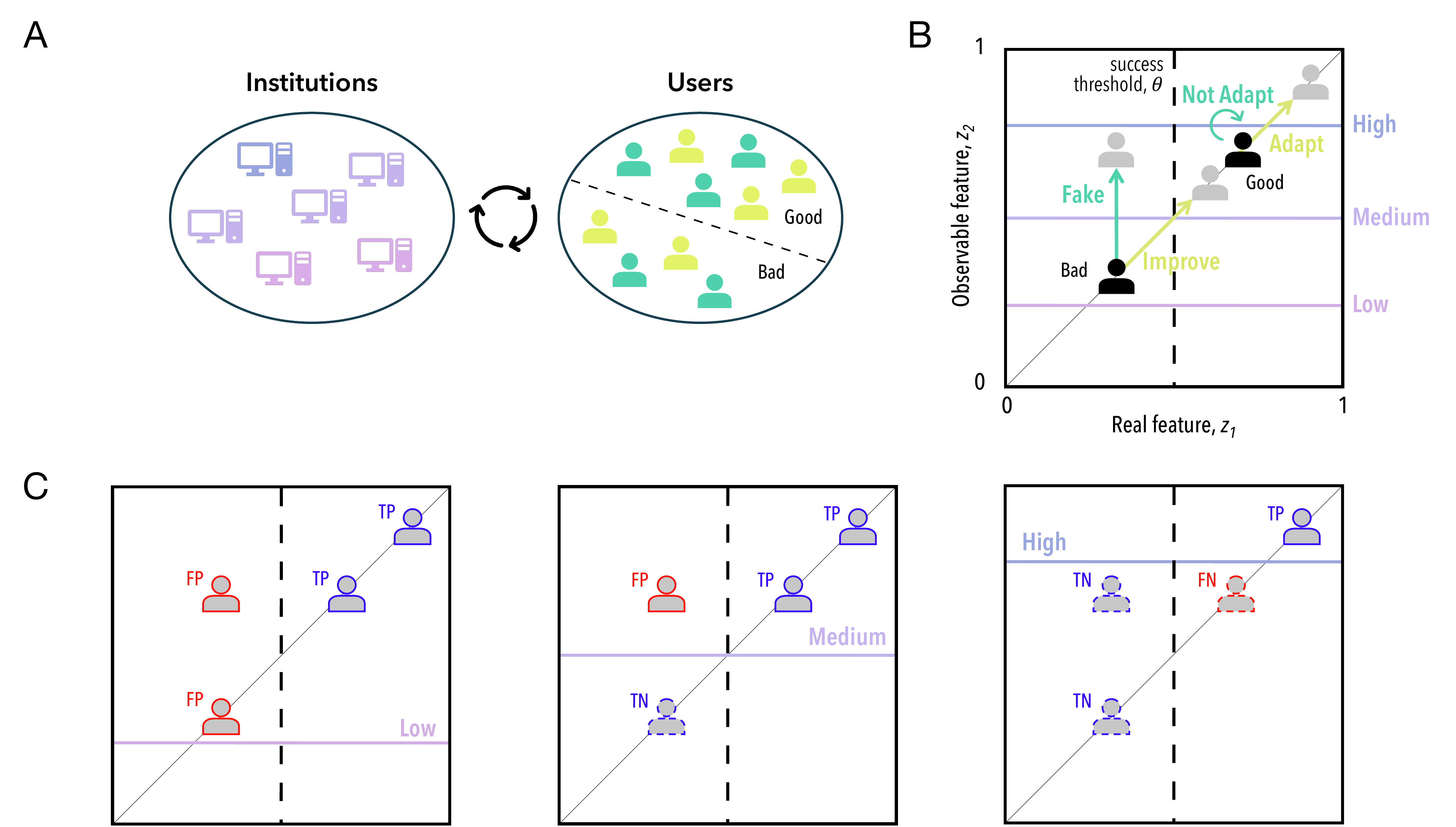}
    \caption{{\textbf{Model illustration.}} (A) Co-adaptation dynamics of strategic classification. We assume that a population of institutions (using classification algorithms) interacts with a population of users (being classified). The population of users is composed of two sub-populations corresponding to Good  and Bad users, respectively, users who have high and low quality. The strategies that each individual can adopt (represented by different colors) co-evolve based on their relative average success. (B) Feature space and strategies. The horizontal axis represents the user’s real feature ($z_1$); the vertical axis represents the user’s feature observable by the institution ($z_2$) and used in the algorithmic classification set by the institution. For simplicity, we depict $1$-dimensional features. 
    We represent the institution's strategies by coloured horizontal lines. They denote the possible locations of the decision boundary deployed by the institution, which can be \textbf{Low}, \textbf{Medium}, or \textbf{High}. 
    The dashed vertical line denotes the success threshold $\theta$ determining a user's true success.
    A Good user has their initial real feature above the success threshold (to the right of the dashed line); therefore, they are a successful candidate. On the contrary, a Bad user has their initial real feature below the success threshold (left of the dashed line); therefore, they are not a successful candidate (unless they honestly improve their initial condition, as we specify below). We represent the users' strategies by coloured arrows, which express a change in the feature space. A Good user can decide whether to \textbf{Adapt} (paying a cost to improve their condition) or \textbf{Not adapt} (remain as they are at no cost). A Bad user can decide to \textbf{Improve} (paying a cost to truthfully improve their initial condition, i.e., to change their real features) or \textbf{Fake} (paying a lower cost to manipulate their initial condition, i.e., to change only their observable features). The black avatars refer to the initial condition ($\tau=0$), whereas the gray avatars refer to the final condition ($\tau=1$), that is, before and after adaptation occurs. (C) Classification outcomes. We show the possible classification outcomes for all institution's strategies: false positive ($FP$) in full red, false negative ($FN$) in dashed red, true positive ($TP$) in full blue, and true negative ($TN$) in dashed blue. In our baseline scenario, a \textbf{Low} institution accepts all users, no matter their strategy (left panel). A \textbf{Medium} institution 
    rejects a Bad user that does not adapt, and accepts all remaining cases (middle panel). A \textbf{High} institution only accepts a Good user who adapts (right panel).}
    \label{fig1}
\end{figure}

In our setup, we allow institutions to decide where their decision boundary lies, from low (lenient or permissive algorithm) to high (strict algorithm).
In turn, users decide how to disclose personal information used by the algorithms.
For example, users can decide how to invest their effort in a loan application, whether paying a high cost for improving their application, a low cost to provide manipulated information, or paying no cost for no feature change.
We ask which of these strategies may evolve and eventually stabilize in the respective populations of users and institutions (Fig.~\ref{fig1}A).
We start by showing that it is hard to achieve moderate institutions --- that are fair to the users --- and, at the same time, high algorithmic performance.
Institutions tend to become strict, and thereby inducing a high social cost.
This problem can be alleviated in two ways.
The first is through the development of manipulation-proof algorithms. 
Yet, achieving full robustness is often impractical due to technical limitations.
An alternative approach is to provide recourse, that is, offering actionable explanations to users on how they can improve in order to be accepted \cite{Ustun:FAT:2019}. 
This scenario also contributes to mitigating undesirable outcomes by fostering a cyclical dynamic between institutions and users: as institutions ease their decision boundaries, users may respond with more faking; in turn, institutions grow stricter, eventually leading to users improvement, supported by recourse mechanisms.

Notably, our model introduces the aspect of frequency-dependent selection, where the proportions of strategies adopted by users and institutions in the co-evolving populations determine which strategies are more fit at any given time.
This approach offers a natural account of the long-run dynamics of strategic classification, paving the way for a clearer understanding of institutions deploying classifiers and users interact, 
particularly taking into account potential existing trade-offs. 
The remainder of the paper is structured as follows.
In Section~\ref{section-model}, we describe our model setup, defining the general game, and introduce a baseline scenario.
In Section~\ref{section-results}, we show the results, introduce new scenarios, and discuss them.
We conclude with Section~\ref{section-conclusion}, where we also discuss limitations of our approach and suggest directions for future research on this field.

\section{Model}
\label{section-model}

\subsection{The general game}

We define an asymmetric game between two types of players, the \textit{institution} and the \textit{user}.
The {institution} aims to accurately classify the {user} in order to provide a service. For example, in the context of banking, the {institution} can be a bank developing an algorithm to decide whether to give a loan or not to a customer. 
In this case, the service is money lending and the algorithm considers information provided by users to predict the client's quality or credit score. 
In the application process, the {user} discloses information about their features (or attributes) to the {institution}. 
We assume that the algorithm deployed by the bank takes as input self-reported information provided via the loan application procedure. 
The bank sets a threshold or boundary on the credit score that determines whether a customer will be classified as positive (i.e., getting the credit) or negative (i.e., not getting the credit).

We assume that an institution may choose to set its score threshold to classify users at three different levels, \textbf{Low} (\textbf{L}), \textbf{Medium} (\textbf{M}) or \textbf{High} (\textbf{H}) (horizontal lines in Fig.~\ref{fig1}B). 
In other words, institutions can be lenient, moderate or harsh, respectively, where a lenient institution accepts lower quality applications from users and a harsh institution only accepts high quality ones.
Formally, we define the institutions' strategy set $\bar{S}_I$ as 
\begin{equation}\label{equation-institution_strategy_set}
  \bar{S}^I~:=~\{{S}^I_i, i=0,1,2\}:=~\{\textbf{Low}, \textbf{Medium}, \textbf{High}\}.
\end{equation}

In turn, users' strategies correspond to adaptations in their features' space, which may affect their score.
Let $z_1(\tau) \in [0,1]$ be a (normalized) true feature value (e.g., real income), and $z_2(\tau) \in [0,1]$ be a (normalized) observable feature value (e.g., declared income) at moment $\tau$, where $\tau\in\{0,1\}$.
We can equate $z_1$ as the true score and $z_2$ as the predicted score by the institution.
In Fig.~\ref{fig1}B, we represent $z_1$ on the x-axis and $z_2$ on the y-axis.
The time $\tau=0$ refers to the initial condition (before actions are taken), and $\tau=1$ refers to the moment after actions are taken.
We assume that initially the observed feature corresponds to the real one, that is, $z_1(0)=z_2(0)$ --- in Fig.~\ref{fig1}B, individuals initially lie on the diagonal.
Only after actions are taken, these quantities may differ. 
Moreover, $z_1(1)\neq z_2(1)$ means that an individual provided erroneous information --- the true and observable features do not match.

Let $\theta$ be the success threshold determining a user's true success or quality (dashed line in Fig.~\ref{fig1}B), such that, for any $\tau$,
\begin{itemize}
    \item if $z_1(\tau)\geq\theta$, the user is successful (e.g. can repay a loan),
    \item if $z_1(\tau)<\theta$, the user is not successful (e.g. is not able to repay a loan).
\end{itemize}

In our approach, it is useful to distinguish {users} in terms of their true quality prior to any action, that is, at $\tau=0$.

We denote the user type by $u$, where $u\in\{\textit{G},\textit{B}\}$. 
Specifically, \begin{itemize}
    \item if $z_1(0)\geq\theta$, the user is denoted by Good ($u=G$), 
    \item if $z_1(0)<\theta$, the user is denoted by Bad ($u=B$). 
\end{itemize}
Hence, in Fig.~\ref{fig1}B, the Bad user is located left of the dashed line, and the Good user is on the right (black avatars).

We assume that {users} can choose from three possible actions when interacting with an {institution} (e.g., when applying for a bank loan).
When providing personal information, {users} can (i) disclose their features as they are (i.e. without any adaptation) at no cost, (ii) fake or misreport their features at a low cost or (iii) improve their features and report them truthfully at a high cost.
In the following, we consider only two of these strategies for each type of {user}.
This allows us to reduce the dimensionality of the problem without impacting the reasoning and the scope of the analysis.
We show in Appendix~\ref{appendix-strategy-sets} how this simplification does not affect our interpretation of the results.

Formally, users can adopt one strategy from the set of strategies $S^G$ and $S^B$ of the Good and Bad user, respectively, where 
\begin{align}
    \label{good_user_strategy_set} S^G~&:=~\{S^G_i, i=1,2\}:=~\{\textbf{Not adapt}, \textbf{Adapt}\},\\
    \label{bad_user_strategy_set} S^B~&:=~\{S^B_i, i=1,2\}:=~\{\textbf{Fake}, \textbf{Improve}\}.    
\end{align}
   
\noindent A Good user, when deciding to \textbf{Not adapt} (\textbf{NA}), keeps their initial score, $z_1(0)=z_1(1)=z_2(1)$, and incurs no cost.
When deciding to \textbf{Adapt} (\textbf{A}), the user increases their true score, $z_1(0)<z_1(1)=z_2(1)$, at some cost.
As for a Bad user, they can choose to \textbf{Fake} (\textbf{F}) or to \textbf{Improve} (\textbf{I}).
To \textbf{Fake} means appearing as having a higher score than they truly have, $z_1(0)=z_1(1)<z_2(1)$, that is, to provide false or manipulated information at a low cost. 
To \textbf{Improve} means to increase one's true score, $z_1(0)<z_1(1)=z_2(1)$ at a high cost.
In Fig.~\ref{fig1}B, the four different actions are depicted by the coloured arrows, which show the variation in the feature space ($z_1$,~$z_2$) from $\tau=0$ to $\tau=1$.

The interaction between the {institutions} and the {users} can be represented by a game defined by the following generic payoff matrices,
\begin{itemize}
    \item $\mathcal{I}^G=(\mathcal{I}^G_{ij})$, the institution payoff matrix against {Good users},
    \item $\mathcal{I}^B=(\mathcal{I}^B_{ij})$, the institution payoff matrix against {Bad users},
    \item $\mathcal{U}^G=(\mathcal{U}^G_{ij})$, the {Good users} payoff matrix,
    \item $\mathcal{U}^B=(\mathcal{U}^B_{ij})$, the {Bad users} payoff matrix, 
\end{itemize}   
where $\mathcal{I}^G_{ij}$ denotes the payoff that the institution obtains when playing strategy $S^I_i$ against a Good user playing strategy $S^G_j$, where $i=0,1,2$ and $j=1,2$.
The other payoff matrices are analogously defined.
We can represent the game by the following tabular scheme
\begin{equation}
	\label{equation-general-matrix}
    \begin{tabular}{ c c || c c | c c }
		& &	\multicolumn{2}{c|}{Good user} 	& 	\multicolumn{2}{c}{Bad user} \\
		& & $\mathbf{\;Not \; adapt\;}$ & $\mathbf{\;Adapt\;}$ & $\mathbf{\;Fake\;}$ & $\mathbf{\;Improve\;}$ \\
		\hline \hline
		\multirow{3}{4em}{Institution} & $\mathbf{\;Low\;}$ & 
        $\mathcal{I}^G_{01}, \mathcal{U}^G_{01}$ & $\mathcal{I}^G_{02}, \mathcal{U}^G_{02}$ & $\mathcal{I}^B_{01}, \mathcal{U}^B_{01}$ & $\mathcal{I}^B_{02}, \mathcal{U}^B_{02}$  \\& $\mathbf{\;Medium\;}$ & 
        $\mathcal{I}^G_{11}, \mathcal{U}^G_{11}$ & $\mathcal{I}^G_{12}, \mathcal{U}^G_{12}$ & $\mathcal{I}^B_{11}, \mathcal{U}^B_{11}$ & $\mathcal{I}^B_{12}, \mathcal{U}^B_{12}$  \\
		& $\mathbf{\;High\;}$ & $\mathcal{I}^G_{21}, \mathcal{U}^G_{21}$ & $\mathcal{I}^G_{22}, \mathcal{U}^G_{22}$ 
        & $\mathcal{I}^B_{21}, \mathcal{U}^B_{21}$ & $\mathcal{I}^B_{22}, \mathcal{U}^B_{22}$  \\
	\end{tabular}
\end{equation}

\noindent The rows correspond to the institution's decisions, and the columns correspond to each user type's decisions. 
By construction, the institution does not know whether it faces a Good or Bad user --- that is what the institution is trying to predict in the first place. 
Thus, the institution's strategy holds for both types of players.
In other words, whatever kind of classification algorithm the institution decides to deploy, it necessarily applies to all users simultaneously.
This setup provides a general formulation of our problem.
Next, we define a particular case.

\subsection{A baseline scenario}
\label{subsection-baseline}

We now build a baseline case of the game introduced in the previous section.
Before assigning payoffs to the game, it is useful to reason in terms of possible classification outcomes (Fig.~\ref{fig1}C).
We start from the point of view of a \textbf{Low} institution.
A \textbf{Low} institution is defined by setting its score threshold so low that it accepts all users (Fig.~\ref{fig1}C left).
Therefore, according to this institution, all cases with $z_1(1)<\theta$ are false positives ($FP$), and all cases with $z_1(1)\geq\theta$ are true positives ($TP$). 
Differently, an institution using a \textbf{Medium} threshold, does not accept an unchanged application from a Bad user, yielding that case a true negative ($TN$) (Fig.~\ref{fig1}C middle).
Finally, a \textbf{High} institution, setting their threshold so high, only accepts Good users that further improve.
This makes a Good user that does not adapt being classified as negative ($FN$).
At the same time, a \textbf{High} institution is able to accurately classify faking behavior ($TN$) (Fig.~\ref{fig1}C right). 
We show in Appendix~\ref{appendix-strategy-sets} that \textbf{Low} is either equivalent to or dominated by other strategies.
Hence, we drop \textbf{Low} from the institution's strategy set in the following without loss of insight.
We summarise the above in the following matrix

\begin{equation}
	\label{outcome-matrix-baseline}
    \begin{tabular}{  c || c c | c c }
		& 	\multicolumn{2}{c|}{Good} 	& 	\multicolumn{2}{c}{Bad} \\
		& $\mathbf{\;Not \; adapt\;}$ & $\mathbf{\;Adapt\;}$ & $\mathbf{\;Fake\;}$ & $\mathbf{\;Improve\;}$ \\
		\hline \hline
		$\mathbf{\;M\;}$ & $TP$ & $TP$ & $FP$ & $TP$  \\
		$\mathbf{\;H\;}$ & $FN$ & $TP$ & $TN$ & $FN$ \\
	\end{tabular}.
\end{equation}

We can now assign actual payoffs to each outcome.
We assume that the {institution} earns a payoff of $\rho$ in the case of a $TP$. 
This parameter denotes the bank's gain from a successful lending operation, for example, linked to the gain from the loan's interest rate.
In the case of a $FP$, we assume that the institution has a loss worth of $\lambda$. 

When a {user} is classified as negative ($N$), i.e., when the institution rejects a user's application, we assume the institution gets a null payoff (the service is not provided).
As for the {user}, they obtain a benefit $b$ when they are accepted (whether they are a $TP$ or a $FP$).
Similar to the institution, we assume that when a {user} is classified as negative they obtain a null payoff.
Moreover, the user pays a cost $c_I$ for adapting or improving and a cost $c_F$ for faking.
Hence, we write the payoff matrices of the game described above as
\begin{equation}
	\label{payoff-matrix-baseline}
    \begin{tabular}{  c || c c | c c }
		& 	\multicolumn{2}{c|}{Good} 	& 	\multicolumn{2}{c}{Bad} \\
		& $\mathbf{\;Not \; adapt\;}$ & $\mathbf{\;Adapt\;}$ & $\mathbf{\;Fake\;}$ & $\mathbf{\;Improve\;}$ \\
		\hline \hline
		$\mathbf{\;M\;}$ & $\rho$, $b$ & $\rho$, $b - c_I$ & $-\lambda$, $b - c_F$ & $\rho$, $b - c_I$  \\
		$\mathbf{\;H\;}$ & $0$, $0$ & $\rho$, $b - c_I$ & $0$, $ - c_F$ & $0$, $ - c_I$ \\
	\end{tabular}
\end{equation}

\noindent where all parameters $\rho$, $\lambda$, $b$, $c_I$ and $c_F$ take real positive values. 
As stated before, we consider the cost of improvement higher than the cost of faking, i.e., $c_I>c_F$. 
We also assume that $b>c_I$,  that is, the cost of improving is never larger than the benefit of receiving a loan for the {user}, irrespective of being Good or Bad.

\subsection{Evolutionary dynamics}

The previous section introduces a baseline game. But how will individuals adapt over time, while interacting along the payoff matrices introduced? Evolutionary game theory provides mathematical tools that allow to anticipate such behavioural dynamics.

Let us consider a population of institutions and two populations of users (Good and Bad) that continually adapt to each other over time.
To model their co-adaptation process, we use the so-called \textit{replicator dynamics}~\cite{Hofbauer:book:1998}.
Essentially, replicator dynamics postulates that the frequency of a strategy in a population increases if and only if it yields a payoff above average.
We can interpret it as more successful strategies being adopted more often and, thus, growing in the population.
Before writing the replicator equations, we introduce some definitions.

We assume that the set of strategies $S^I$ available to the institution is a subset of $\bar{S}^I$ (Eq.~(\ref{equation-institution_strategy_set})), $S^I=\{S^I_i,i=1,2\}\subset \bar{S}^I$, as we have droped the \textbf{Low} strategy.
We define the states of the populations at time $t$:
\begin{itemize}
    \item The state of the sub-population of {institutions} at time $t$ is defined by the vector $\textbf{x}(t)=(x_1(t),x_2(t))$, where $x_i(t)$ is the fraction of institutions using the strategy $S^I_i\in~S^I$, with $x_1(t)+x_2(t)=1, \forall t$. 

    \item The state of the sub-population of {Good users} at time $t$ is defined by the vector $\textbf{y}^G(t)=(y^G_{1}(t), y^G_{2}(t))$, where $y^G_j(t)$ is the the fraction of {Good users} using the strategy $S^G_j\in S_G$, with $y^G_{1}(t) + y^G_{2}(t) = 1, \forall t$.

    \item The state of the sub-population of {Bad users} at time $t$ is defined by the vector $\textbf{y}^B(t)=(y^B_{1}(t), y^B_{2}(t))$, where $y^B_j(t)$ is the fraction of {Bad users} using the strategy $S^B_j\in~S^B$, with $y^B_{1}(t) + y^B_{2}(t) = 1, \forall t$.

    \item Finally, the state of the entire population --- institutions and users --- at time $t$ is fully defined by $\textbf{X}(t)=(x_1(t),{y}^G_{1}(t),{y}^B_{1}(t))$.
\end{itemize}
In the following, we mostly use $\textbf{X}$ so as to refer to a state of the whole system.

We assume infinite and well-mixed populations of players. 
Below, the notation $(\mathcal{A})_i$ denotes the $i$-th row of matrix $\mathcal{A}$.
Let $f^I_{i}$ denote the average payoff or fitness obtained by an institution when implementing strategy $S^I_i$. We can write it as 
\begin{equation}
 \begin{split}
      f^I_{i}&= p_G \; (\mathcal{I}^G)_i \cdot (\textbf{y}^G)^\top + p_B \; (\mathcal{I}^B)_i \cdot (\textbf{y}^B)^\top\\
      p_B&=1-p_G,
 \end{split}
\end{equation}
where $p_G \in [0,1]$ and $p_B \in [0,1]$ represent, respectively, the proportion of Good and Bad users w.r.t. the population of users; $(\mathcal{I}^G)_i$ and $ (\mathcal{I}^B)_i$ denote, respectively, the $i$-th rows of matrices $(\mathcal{I}^G)$ and $(\mathcal{I}^B)$, introduced in Eq.~\eqref{equation-general-matrix} and $\textbf{v}^\top$ the transposed of vector $\textbf{v}$.

Let $\bar{f}^I$ denote the average payoff of the institutions' population. It is given by
\begin{equation}
 \bar{f}^I= \textbf{f\,}^I \cdot \textbf{x}^\top
\end{equation}
where $\textbf{f\,}^I = (f^I_{1},f^I_{2})$.\\

Similarly, let $f^u_{j}$ denote the average payoff obtained by a $u$-type user implementing strategy $S^u_j$, where $u = G, B$. It is given by 
\begin{equation}
 f^u_{j} = (\mathcal{U}^u)_j \cdot \textbf{x}^\top
\end{equation}
Finally, let $\bar{f}^u$ denote the average payoff or fitness of the type-u users' population. It is given by
\begin{equation}
 \bar{f}^u= \textbf{f\,}^u \cdot (\textbf{y}^u)^\top
\end{equation}
where $\textbf{f\,}^u = (f^u_{1}, f^u_{2})$.

The evolutionary dynamics is defined by the continuous time replicator dynamics \cite{Hofbauer:book:1998}, characterized by the following differential equations 
\begin{align}
\dot{x}_i &= r \cdot x_i(f^I_{i}-\bar{f}^I), \mbox{ with } i=1,2 \\
\dot{y}_j^u &= y_j^u(f^u_{j}-\bar{f}^u), \mbox{ with } j=1,2 \mbox{ and } u\in \{\textbf{G},\textbf{B}\}
\end{align}

\noindent where the dot symbol on the l.h.s. denotes the time derivative and $r \in \mathbb{R}^+$ is a constant that determines the evolutionary rate of the institutions compared to that of users.

For example, if $r=2$, {institutions} adapt two times faster than users.\\

Given that we can directly obtain $x_2$ and $y_2^u$ from $1-x_1$ and $1-y_1^u$, respectively, we are left with only three equations.
The replicator equations of the previously defined game in Eq.~\eqref{payoff-matrix-baseline} thus become simply
\begin{equation}\label{equation-replicator-baseline}
 \begin{split}
      \dot{x}_1 &= r \; x_1 \big(1-x_1\big)\big(\rho - \rho \, p_G \, (1-y_1^G) - (\lambda+\rho)(1-p_G) \, y_1^B  \big),  \\
\dot{y}_1^G &= y_1^G \big(1-y_1^G\big) \big(c_I - b(1-x_1)\big), \\
\dot{y}_1^B &= y_1^B \big(1-y_1^B\big) \big(c_I - c_F\big).
 \end{split}
\end{equation}

\noindent We study the properties of this system in the next section.

\section{Results and discussion}
\label{section-results}

We start by presenting the results for the baseline scenario introduced in section~\ref{subsection-baseline}. 
We then explore different scenarios by changing this baseline case.

\subsection{Baseline: Imperfect classifier}
\label{baseline:imp_class}
 
\begin{figure}[h!]
    \centering
    \includegraphics[width=1\linewidth]{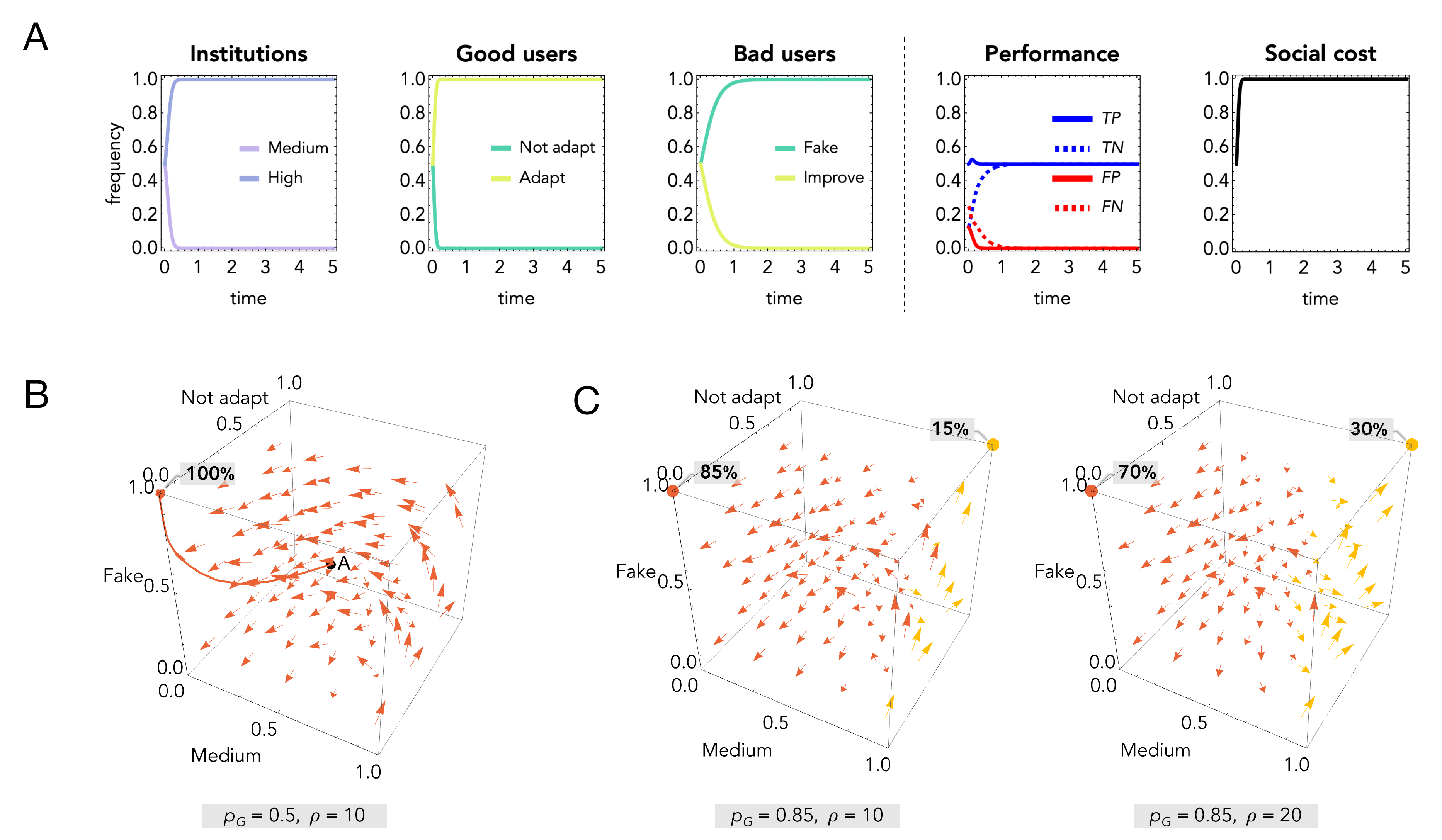}
    \caption{{\textbf{Imperfect classifier.}} Results from replicator dynamics. (A) We numerically integrate Eq.~(\ref{equation-replicator-baseline}) with initial condition $(0.5,0.5,0.5)$. The three left panels show the strategy frequency evolution over time in each population --- $(x_1(t),x_2(t))$ for \textit{Institutions}, $(y^G_{1}(t), y^G_{2}(t))$ for \textit{Good users}, and $(y^B_{1}(t), y^B_{2}(t))$ for \textit{Bad users}. The two panels on the right show the evolution of algorithmic \textit{performance} and the \textit{social cost}. We define performance in terms of the frequencies of \textit{TP}, \textit{TN}, \textit{FP}, and \textit{FN} cases. As an example, we calculate the fraction of \textit{TN} cases by $ x_2 \, (1-p_G) \, y^B_1$, because a \textit{TN} can only occur, in this scenario, when a \textbf{High} institution meets a Bad user who plays \textbf{Fake} (matrix~\ref{outcome-matrix-baseline}). Social cost is defined by the fraction of \textbf{Good} users who \textbf{Adapt}. In terms of behavioural strategies, the system converges to the monomorphic state $(\textbf{H},\textbf{A},\textbf{F})$. Performance converges to half \textit{TP} and half \textit{TN} cases. Social cost reaches the highest level. (B) The 3-dimensional vector field of the dynamical system. The axes correspond to the fractions of institutions playing \textbf{Medium}, Good users playing \textbf{Not adapt}, and Bad users playing \textbf{Fake}. The vector field indicates, for any given point of the state space, the direction of a trajectory starting at that point. Additionally, we plot the trajectory shown in panel (A), denoting by a black circle the initial point and by a red circle the final point. The gray tag pointing at state $(\textbf{H},\textbf{A},\textbf{F})$ shows its respective basin of attraction size, that is, the percentage of trajectories ending at that state. To calculate it, we solve the equation for $8000$ initial conditions defined by a grid with equally spaced points over the state space. (C) Vector fields for the condition $p_G>p_G^*$, in particular $p_G=0.85$, and two values of the parameter $\rho$: $\rho=10$ on the left and center, and $\rho=20$ on the right. The gray tags show the relative size of the basins of attraction, as in panel (B). Here, there are two convergent points, $(\textbf{H},\textbf{A},\textbf{F})$ in red, and $(\textbf{M},\textbf{NA},\textbf{F})$ in yellow. For all panels, the parameters are $\lambda=50$, $\rho=10$, $b=50$, $c_F=1$, $c_I=5$, $p_G=0.5$, $r=1$ (except when stated otherwise).}
    \label{fig2}
\end{figure}

We first focus on the baseline scenario given by the game with payoff matrix in Eq.\eqref{payoff-matrix-baseline}.
We highlight that (i) moderate institutions deploy algorithms unable to detect gaming from users and (ii) harsh institutions, while rejecting users who fake, also reject Good users who do not artificially inflate their features (Fig.~\ref{fig1}C).
We refer to this scenario as ``imperfect classifier'' as moderate institution cannot perfectly classify the users.
We study the dynamics of this interaction through the replicator dynamics.
We start by numerically solving the dynamical system given in Eqs.~\eqref{equation-replicator-baseline} and we plot the frequency of strategies adopted over time.
Starting from the state where the populations have the same proportion of each strategy, $(0.5,0.5,0.5)$, the system eventually reaches the state $(0,0,1)$ (Fig.~\ref{fig2}A, three leftmost panels), that is, where all institutions play \textbf{High}, all Good users \textbf{Adapt} and all Bad users \textbf{Fake}. 

For convenience, when referring to a monomorphic state, we label it using the acronym of the single strategy present in each of the populations.
Therefore, $(0,0,1)$ can also be denoted by $(\textbf{H},\textbf{A},\textbf{F})$.
We also plot the institution's performance in terms of the frequency of each classification outcome --- $TP$, $TN$, $FP$ and $FN$ (Fig.~\ref{fig2}A, fourth panel).
The dynamics converges to a state where half of the cases are true positives and half are true negatives.
Although, in terms of performance, this may appear as a good outcome, the state $(\textbf{H},\textbf{A},\textbf{F})$ is highly undesirable.
First, Bad users do not have an incentive to improve.
Second, as harsh institutions take over, there is a pressure for already Good candidates to incur extra costs in order to appear reliable.
This additional cost of adaptation was previously studied in the strategic classification literature and referred to as ``social cost'' or social burden \cite{Milli:FAT:2019}.
Here, we define it as the fraction of Good users who use the strategy \textbf{Adapt} (i.e., it equates to the light green line in the Good users panel).
For better comparison with institutional performance, we plot the social cost in a separate panel (Fig.~\ref{fig2}A, rightmost panel).

To verify whether other initial conditions lead to the same result, we plot the vector field of the dynamical system (Fig.~\ref{fig2}B).
The vector field indicates the direction of a trajectory starting at that point in the full 3-dimensional state space.
The color red depicts the initial conditions that end in state $(\textbf{H},\textbf{A},\textbf{F})$, shown by a circle of the same color.
We show that all initial conditions considered lead to the undesirable state.

From Eqs.~\eqref{equation-replicator-baseline}, we can also analytically derive the fixed points and the respective linear stability properties \cite{Strogatz:book:1994}.
This allows for a more systematic exploration of the model parameters.
Trivially, all corner states ---  where each of the three populations is monomorphic --- are fixed points of the dynamical system.
However, only the state $(\textbf{H},\textbf{A},\textbf{F})$ is locally stable for all the parameter values we consider.
We provide more details of the stability analysis in Appendix~\ref{appendix-stability}.
Moreover, the state $(\textbf{M},\textbf{NA},\textbf{F})$ is stable as well if 
\begin{equation}\label{equation-fixed-point-condition}
    p_G> p_G^* \equiv \frac{\lambda}{\lambda+\rho} = \frac{1}{1+\rho/\lambda}.
\end{equation}

\noindent Hence, the $(\textbf{M},\textbf{NA},\textbf{F})$ state can be an endpoint of the dynamics provided the proportion of Good users in the population $p_G$ or the ${\rho}/{\lambda}$ ratio is large enough.
The lower $\lambda$, the lower the risk associated with false positives for {\textbf{Medium}} institutions (recall that this corresponds to the loss associated with a failed repayment);
therefore, it is sensible that lower values of $\lambda$ lead to easier conditions for stable {\textbf{Medium}} institutions.
A high gain from true positives, i.e. high $\rho$, has a similar effect.
Even if that means that {\textbf{High}} institutions would also have higher gains.

We note that local linear stability does not imply non-local convergence.
Therefore, it is useful to look at the sizes of the basins of attraction of the several stable fixed points.
A basin of attraction corresponds to the set of initial conditions that lead to a particular ending state.
We can thus ask whether the parameters above also have an effect on the basins of attraction of the two stable points.
To visualize those areas, we plot the vector fields for cases where $p_G>p_G^*$ in Eq.~\eqref{equation-fixed-point-condition} (Fig.~\ref{fig2}C).
We verify that by increasing $\rho$ from $10$ to $20$ (with fixed $\lambda=50$), the basin of attraction of $(\textbf{M},\textbf{NA},\textbf{F})$ increases from $15\%$ to $30\%$.

The state $(\textbf{M}, \textbf{NA}, \textbf{F})$ is a more desirable state to the users, because {\textbf{Medium}} institutions thrive and therefore, Good users do not need to {\textbf{Adapt}} (the social cost is reduced).
However, Bad users fully fake, making this state less attractive to the institutions, which misclassify such cases ($FP$).
Interestingly, the higher the ${\rho}/{\lambda}$ ratio --- which describes the relative size of the institution's gain and loss associated with the loan  --- the more salient the state $(\textbf{M},\textbf{NA},\textbf{F})$ is.

This {\textit{baseline scenario}} captures and highlights the risks of not using strategy-robust algorithms.
It shows that the system either leads to a harmful situation for the users --- Good users suffering from high social costs when institutions are harsh --- or for the institutions, which when moderate become unprotected. 
Thus, this scenario reveals negative impacts on both accuracy and fairness of the classification algorithm.

\subsection{Manipulation-proof classifier}
\label{manip_class}

We now consider a scenario where {\textbf{Medium}} institutions are {\textit{robust}} against faking behaviour. 
For that, we simply change one entry of the previous matrix in Eq.~\eqref{outcome-matrix-baseline}.
We do so by assuming that when a {\textbf{Medium}} institution meets a faking (Bad) user, it classifies them as negative, yielding a true negative ($TN$).
The new outcome matrix becomes
\begin{equation}
	\label{}
    \begin{tabular}{  c || c c | c c }
		& 	\multicolumn{2}{c|}{Good} 	& 	\multicolumn{2}{c}{Bad} \\
		& $\mathbf{\;Not \; adapt\;}$ & $\mathbf{\;Adapt\;}$ & $\mathbf{\;Fake\;}$ & $\mathbf{\;Improve\;}$ \\
		\hline \hline
		$\mathbf{\;M\;}$ & $TP$ & $TP$ & $\textbf{\textit{TN}}$ & $TP$  \\
		$\mathbf{\;H\;}$ & $FN$ & $TP$ & $TN$ & $FN$ \\
	\end{tabular}
\end{equation}
\noindent where we highlight in bold the only difference to the baseline scenario. Similarly, in terms of payoffs, we write 
\begin{equation}
	\label{}
    \begin{tabular}{  c || c c | c c }
		& 	\multicolumn{2}{c|}{Good} 	& 	\multicolumn{2}{c}{Bad} \\
		& $\mathbf{\;Not \; adapt\;}$ & $\mathbf{\;Adapt\;}$ & $\mathbf{\;Fake\;}$ & $\mathbf{\;Improve\;}$ \\
		\hline \hline
		$\mathbf{\;M\;}$ & $\rho$, $b$ & $\rho$, $b - c_I$ & $0$, $ - c_F$ & $\rho$, $b - c_I$  \\
		$\mathbf{\;H\;}$ & $0$, $0$ & $\rho$, $b - c_I$ & $0$, $ - c_F$ & $0$, $ - c_I$ \\
	\end{tabular}.
\end{equation}

\begin{figure}[h!]
    \centering
    \includegraphics[width=1\linewidth]{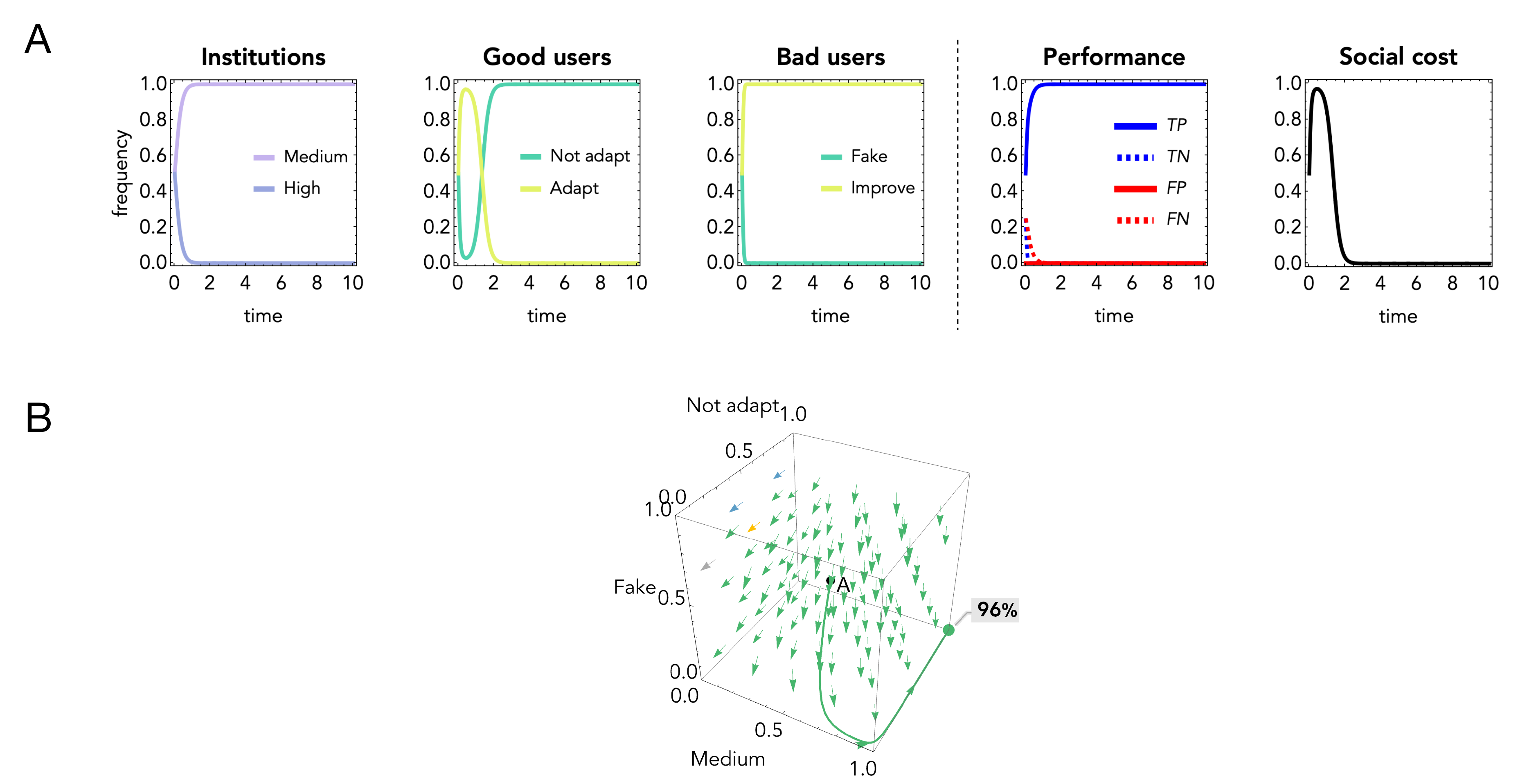}
    \caption{{\textbf{Manipulation-proof classifier.}} This figure is mostly analogous to Fig.~\ref{fig2}. Here, we represent the ideal scenario of deploying classifiers which can accurately identify users faking. Although we observe a transient state where social cost increases, the system converges towards a state where both institutions' performance and users' utility is high. (A) We numerically integrate the replicator equation Eq.~\eqref{equation-replicator-robust} from initial condition $(0.5,0.5,0.5)$. The three panels on the left show the strategy frequency evolution over time, for institutions, Good users and Bad users, respectively. The two panels on the right show the evolution of algorithmic performance, and the social cost. (B) The 3-dimensional vector field of the dynamical system and the trajectory shown in panel (A), denoting by a black circle the initial point and by a green circle the final point. The gray tag pointing at state $(\textbf{M},\textbf{NA},\textbf{I})$ shows its respective basin of attraction size. We calculate it as described in Fig.~\ref{fig2}. For all panels, the parameters are $\lambda=50$, $\rho=10$, $b=50$, $c_F=1$, $c_I=5$, $p_G=0.5$, $r=1$.}
    \label{fig3}
\end{figure}

The replicator equations for this case become
\begin{equation}\label{equation-replicator-robust}
 \begin{split}
      \dot{x}_1 &= r \; \rho \; x_1 \big(1-x_1\big)\big(1 - p_G\,(1-y_1^G) - y_1^B\,(1-p_G)  \big),  \\
\dot{y}_1^G &= y_1^G \big(1-y_1^G\big) \big(c_I - b \, (1-x_1)\big), \\
\dot{y}_1^B &= y_1^B \big(1-y_1^B\big) \big(c_I - c_F - b \, x_1\big).
 \end{split}
\end{equation}

Similarly to the previous scenario, we plot how the strategy frequency changes over time when we start from state $(0.5,0.5,0.5)$ (Fig.~\ref{fig3}A).

We verify that {\textbf{Medium}} institutions grow, as they no longer misclassify the Bad users.
In addition, while {\textbf{High}} institutions are still present in the population, Good users {\textbf{Adapt}}.
As soon as {\textbf{High}} institutions are eliminated, Good users can reduce adaptation.
At the same time, since \textbf{Medium} institutions are now robust, the Bad users have an incentive to \textbf{Improve} --- and be accepted --- instead of \textbf{Fake}.
As a result, the system eventually stabilizes in the state $(\textbf{M},\textbf{NA},\textbf{I})$, which is highly desirable both for users and institutions, with low social cost and high algorithmic performance.

We also verify that almost all initial conditions ($96\%$) lead to the state $(\textbf{M},\textbf{NA},\textbf{I})$, denoted with green color in the vector field (Fig.~\ref{fig3}B).
We show that this is the only linearly stable fixed point of the system for our parameters range (Appendix~\ref{appendix-stability}).

The {\textit{manipulation-proof classifier scenario}} shows how introducing strategy-robust algorithms completely changes the evolutionary dynamics, when compared to the {\textit{baseline scenario}} with no strategy-robust classifier. 
However, one could argue that supposing that \textbf{Medium} institutions have perfectly accurate classifiers is a rather strong assumption.
Therefore, in the following, we consider a scenario that does not rely on this assumption; instead, it includes algorithmic recourse for {\textbf{High}} institutions.

\subsection{Algorithmic recourse}
\label{algo_rec}

Here, we assume that \textbf{Medium} institutions are imperfect classifiers as in the baseline scenario.
However, we introduce the concept of algorithmic recourse in \textbf{High} institutions. 
We have been implicitly considering that, when users adapt, they do not know exactly how to modify certain features to get accepted. 
For concrete examples, they may not know the kind of institution they face --- whether more or less harsh --- or which features are more relevant to modify.
In some real applications, institutions can provide (actionable) recourse \cite{Ustun:FAT:2019}.
In practice, this means that institutions give some explanation to a user who has not been accepted in a first application about how they could have been accepted (i.e., counterfactual explanations). 
With such information, the user can then improve and reapply --- this time, with higher chances of being accepted and, thus, classified as a true positive $TP$.
To capture this possibility in our setup, we rewrite the baseline matrix as
\begin{equation}
	\label{}
    \begin{tabular}{  c || c c | c c }
		& 	\multicolumn{2}{c|}{Good} 	& 	\multicolumn{2}{c}{Bad} \\
		& $\mathbf{\;Not \; adapt\;}$ & $\mathbf{\;Adapt\;}$ & $\mathbf{\;Fake\;}$ & $\mathbf{\;Improve\;}$ \\
		\hline \hline
		$\mathbf{\;M\;}$ & $TP$ & $TP$ & $FP$ & $TP$  \\
		$\mathbf{\;H\;}$ & $FN$ & $TP$ & $TN$ & $\textbf{\textit{TP}}$ \\
	\end{tabular}
\end{equation}
\noindent where we change the entry corresponding to (\textbf{H}, \textbf{I}) from  a \textit{FN} to a \textit{TP}, as highlighted in bold.
The respective payoff matrix is
\begin{equation}
	\label{}
    \begin{tabular}{  c || c c | c c }
		& 	\multicolumn{2}{c|}{Good} 	& 	\multicolumn{2}{c}{Bad} \\
		& $\mathbf{\;Not \; adapt\;}$ & $\mathbf{\;Adapt\;}$ & $\mathbf{\;Fake\;}$ & $\mathbf{\;Improve\;}$ \\
		\hline \hline
		$\mathbf{\;M\;}$ & $\rho$, $b$ & $\rho$, $b - c_I$ & $-\lambda$, $b - c_F$ & $\rho$, $b - c_I$  \\
		$\mathbf{\;H\;}$ & $0$, $0$ & $\rho$, $b - c_I$ & $0$, $ - c_F$ & $\rho$, $b - c_I$ \\
	\end{tabular}
\end{equation}
We note that, in the case of actionable recourse, there might still be room for manipulation. 
Therefore, we assume that {\textbf{High}} institutions also have in place detection mechanisms to spot cheating behaviour after actionable recourse is provided. 
In this sense, we maintain that {\textbf{High}} institutions classify faking Bad users as negative ($TN$), excluding overlooked cheating.

The replicator equations for this scenario stand as 
\begin{equation}\label{equation-replicator-recourse}
 \begin{split}
\dot{x}_1 &= r \; x_1 \big(1-x_1\big) \big(\rho \, p_G\,y_1^G - \lambda \,(1-p_G) \, y_1^B  \big),  \\
\dot{y}_1^G &= y_1^G \big(1-y_1^G\big) \big(c_I - b \, (1-x_1)\big), \\
\dot{y}_1^B &= y_1^B \big(1-y_1^B\big) \big(c_I - c_F - b \, (1-x_1)\big).
 \end{split}
\end{equation}

\begin{figure}[h!]
    \centering
    \includegraphics[width=1\linewidth]{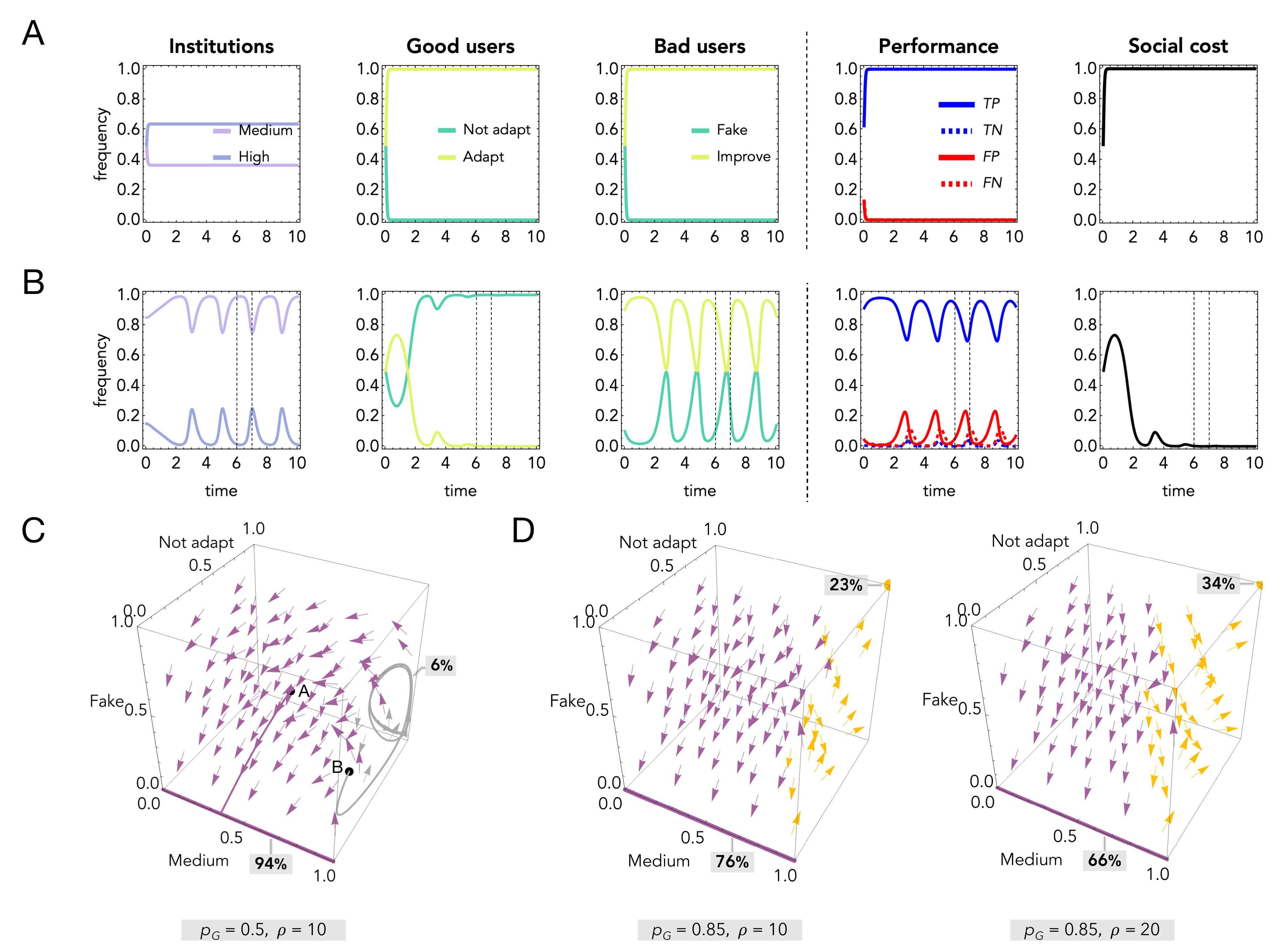}
    \caption{{\textbf{Algorithmic recourse}}. If Institutions allow even Bad users to improve and be thereby being classified as true positives (i.e., providing algorithmic recourse), we observe that {\textbf{Medium}} and {\textbf{High}} institutions co-exist, leading to high improvement rates from Bad individuals (which is desirable) yet high social cost (undesirable). Results from replicator dynamics. All panels are analogous to previous figures. For (A), we start with the initial condition $(0.5,0.5,0.5)$; for (B), we start with the initial condition $(0.85,0.5,0.1)$. The vertical dashed lines delimit the time interval between $6$ and $7$ only to help guide the eye through the cycle along all three dimensions. (C) The 3-dimensional vector field of the dynamical system and the trajectories shown in panels (A) and (B), denoting the initial points by black circles. (D) Vector fields for the condition $p_G>p_G^*$, in particular $p_G=0.85$, and two values of the parameter $\rho$: $\rho=10$ on the left, and $\rho=20$ on the right. The gray tags show the basins of attraction of $(x_1,\textbf{A},\textbf{I})$ (in purple) and $(\textbf{M},\textbf{NA},\textbf{F})$ (in yellow). We calculate them as described in Fig.~\ref{fig2}. For all panels, the parameters are $\lambda=50$, $\rho=10$, $b=50$, $c_F=1$, $c_I=5$, $p_G=0.5$, $r=1$ (except when stated otherwise).}  
    \label{fig4}
\end{figure}

Starting from state $(0.5,0.5,0.5)$, where the populations have the same fraction of each strategy, both \textbf{Medium} and \textbf{High} institutions end up coexisting (Fig.~\ref{fig4}A).
In particular, for this initial condition, \textbf{High} institutions reach a proportion of about $65\%$.
As before, this forces Good users to \textbf{Adapt}.
On a positive note, Bad users {Improve} --- that way, they are accepted by all institutions, including the \textbf{High} ones.
This shows that algorithmic recourse can incentivize full improvement.
Yet, it does not prevent the social cost on the Good users (Fig.~\ref{fig4}A right).

Interestingly, when starting with more \textbf{Medium} institutions and improving users, we observe a cyclic dynamics (Fig.~\ref{fig4}B).
The cycle occurs among the institutions and the Bad users' populations, while the Good users reach full non-adaptation after some initial adjustment.
As institutions are close to being fully \textbf{Medium}, Bad users drop improvement and start faking more, as they are still accepted but at a lower cost. 
Institutions, being cheated upon, start being stricter (the dashed lines Fig.~\ref{fig4}B delimiting times $6$ to $7$ help guiding the eye).
In turn, because Bad users are only accepted by \textbf{High} institutions if they \textbf{Improve}, they start improving again.
However, institutions cannot grow stricter indefinitely, as they would miss opportunities with Good users, since the latter are locked in \textbf{Not adapt} (and thus, rejected by \textbf{High} institutions).
The pattern repeats then.

We show the two types of trajectory on the 3-dimensional vector field in Fig.~\ref{fig4}C.
We verify that most trajectories end up in the full adaptation and full improvement edge (in purple), that is, in states $(x_1,\textbf{A},\textbf{I})$ with $0<x_1<1$.
The cycles (in gray) are confined to a relatively high fraction of {\textbf{Medium}} and \textbf{Improve} strategies. 
Hence, this case leads to more favourable states compared to both the non-cycling trajectories and the baseline scenario. 
Importantly, the cycles exist even when $p_G$ is not too high.
In fact, for the cycles to appear, $p_G$ must be sufficiently small, namely $p_G < p_G^*$, where $p_G^* = \frac{\lambda}{\lambda+\rho}$, as previously defined in Eq.~\eqref{equation-fixed-point-condition}.
In that case, there is a stable fixed point at $(\frac{b+c_F-c_I}{b},1,\frac{\rho \, p_G}{\lambda \, (1-p_G)})$, which corresponds to the centre of the cycles.
We further explore the cyclic dynamics in Appendix~\ref{appendix-cycles} (Fig.~\ref{fig-si1}).

For the complementary condition, $p_G > p_G^*$, the state $(\textbf{M},\textbf{NA},\textbf{F})$ is stable, as in the baseline scenario. 
We show this regime in Fig.~\ref{fig4}D, for two different values of $\rho$.
Similar to the baseline scenario, here, increasing $\rho$ (for fixed $\lambda$) also increases the size of the basin of attraction of the state $(\textbf{M},\textbf{NA},\textbf{F})$.

This scenario shows that recourse helps achieve relatively advantageous states for all stakeholders, even for intermediate proportions of Good users, $p_G$.
However, this occurs only for a limited set of initial conditions (in particular, sufficiently high fractions of \textbf{Medium} and \textbf{Improve} strategies), and in a cyclic dynamics.

\subsection{Effect of re-training rate}

Finally, we explore the effect of $r$, the institutions' adaptation rate.
This parameter determines how fast the institutions adapt their strategies in relation to the users --- in practice, defining the relative speed of users modifying their behaviours and re-training algorithms.  

\begin{figure}[h!]
    \centering   \includegraphics[width=1\linewidth]{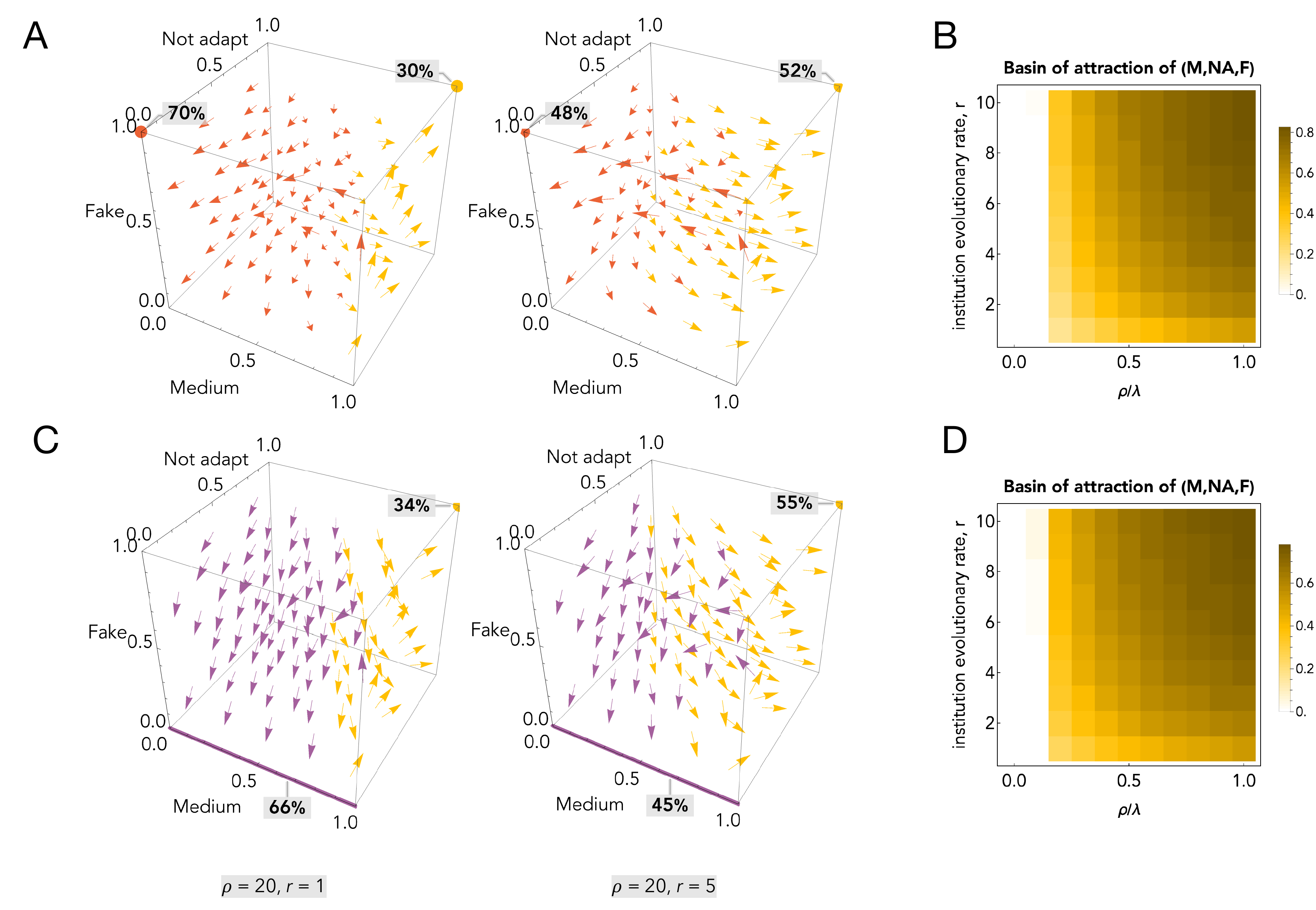}
    \caption{\textbf{Effect of parameters $r$ and $\rho$. (A,B) Imperfect classifier scenario and (C,D) Algorithmic recourse scenario.} Parameter $\rho$ represents the payoff of the institution when accurately identifying a positive example (True Positive, $TP$). Parameter $r$ captures the relative speed at which the institution reacts compared to the users. (A,C) Vector fields for $r=1$ (left) and $r=5$ (right). (B,D) The plots show the basin of attraction size of state $(M, NA, F)$., where Institutions remain Medium and Bad individuals Fake, enabling several $FP$ cases. We vary the ratio $\rho/\lambda$ (x-axis) and varying adaptation rate $r$ (y-axis). We observe that adapting faster (high $r$) can be counterproductive by enlarging the basing of attraction towards $(M, NA, F)$. For all panels, the parameters are $\lambda=50$, $\rho=20$, $b=50$, $c_F=1$, $c_I=5$, $p_G=0.85$ (except when stated otherwise).}
    \label{fig5}
\end{figure}

Acting as a scaling factor on the replicator equations, and as seen in the above analysis, $r$ has no effect on the system's fixed points --- that is, it does not determine their location or stability.
However, we show that it has an important impact on the size of the basins of attraction.
For example, in the baseline scenario (Fig.~\ref{fig5}A), the basin of attraction of state $(\textbf{M},\textbf{NA},\textbf{F})$ increases from $30\%$ to $52\%$, when $r$ increases from $1$ to $5$.
This is at the expense of the basin of attraction of $(\textbf{H},\textbf{A},\textbf{F})$ decreasing with $r$.

In Fig.~\ref{fig5}B, we summarise the effect of both $r$ and $\rho/\lambda$ on the basin of attraction of $(\textbf{M},\textbf{NA},\textbf{F})$.
For the algorithmic recourse scenario, the effect is similar (Fig.~\ref{fig5}C,D). 
For both scenarios, the higher $\rho/\lambda$ and $r$, the larger is the basin of attraction of $(\textbf{M},\textbf{NA},\textbf{F})$.
Thus, when institutions have a higher gain-to-loss ratio or change strategies faster, they more often end up in their least preferred outcome.
Remind that at $(\textbf{M},\textbf{NA},\textbf{F})$, the institutions become unprotected against faking behavior.
Somewhat counterintuitively, this implies that when taking into account the user-algorithm feedback loop, that neither offering loans at higher interest rates or re-adapting quickly would not necessarily benefit the institution.

\section{Conclusion}
\label{section-conclusion}

This paper introduces a framework to study the co-evolving dynamics of strategic classification between institutions and users. 
Typically, strategic classification considers one-shot interaction between a person and an algorithm, often modelled as a Stackelberg game \cite{Hardt:ITCS:2016}.

Most previous research has focused on how to make algorithms robust against potential user manipulation.
Instead, our goal is to understand the collective dynamics between users and institutions from a complex system perspective.
To that end, we resort to evolutionary game theory, a framework that combines game theory and population dynamics. 
We show that this approach not only sharpens the intuitive understanding of the strategic classification problem but also yields some new, unexpected results.

Taking as inspiration the domain of algorithms applied in credit lending, we model populations of users (i.e., customers applying to credit) and institutions (i.e., banks deploying an algorithm) that continuously adapt to each other over time.
We ask which strategies each individual adopts in the long run.
We assume that institutions may vary in how lenient they are with respect to their decision boundary, i.e., the score threshold for granting credit to a client.
In turn, users can invest in their application by improving their quality, manipulating the information provided at a lower cost, or by making no investment at all.
We consider several scenarios, each representing different interaction paradigms.

Our baseline scenario models the case where moderate institutions are not robust against manipulation.
In that case, we show that the system mostly converges towards institutions setting a high boundary, and users either paying excessive costs to meet the institutions' expectations (if they are Good-quality users) or providing fake information (if they are Bad-quality users).
A different endpoint --- where institutions deploy an intermediate threshold, and hence, Good users do not need to incur extra costs, but Bad users cheat anyway --- is possible only if (i) there are sufficiently many Good users in the population or (ii) the gain of the institution from a successful loan-giving relative to the loss from an unsuccessful case is sufficiently high.
This scenario illustrates the problem of strategic classification. 
The two possible solutions create trade-offs for different stakeholders --- one favours high algorithmic performance at the cost of increased social burden, while the other reduces social costs but compromises performance.

We can overcome the above dilemma when considering that all institutions have in place classifiers that are robust against cheating (i.e., strategy-robust algorithms).
In this case, institutions can afford to be more moderate, which, in turn, allows Good users to be accepted without extra costs, and Bad users are required to improve honestly.
However, this would require that even moderate algorithms can perfectly distinguish Good from Bad users, which, in practice, may look like a very strong assumption to be implemented for all institutions and all users.  
As such, we explore a third scenario that does not require the above assumption of perfect classification.
Instead, we consider that strict institutions (i.e., the ones setting a high threshold for classification) can provide {\textit{actionable recourse}} to their unsuccessful users, meaning those who have been classified as negative, thus rejected for the loan.
Interestingly, in this case, we can observe cycling dynamics. 
The proportions of moderate institutions and of honest users in the populations oscillate between intermediate to high levels.
This is a more desirable outcome, as it creates less negative impacts on all populations, alleviating the social cost on users while keeping algorithmic performance relatively high.

Notably, we also explore the effect of the relative speed at which institutions and users update their strategies.
We show that the faster the institutions re-adapt to the users' populations, the more likely the institutions end up in their least preferred outcome.
This observation has interesting connections to the work by Zrnic et al. (2021) \cite{Zrnic:ANeurIPS:2021}.
There, the authors consider a decision-maker (akin to our institution) interacting with a population of users who learn and adapt dynamically over time.
Importantly, they make the learning rates (or update frequency) a tunable variable, in sharp contrast with previous works where users instantly best-respond to a known algorithm deployed by the decision-maker.
They find that when the decision-maker updates faster, there is a lower risk for both players.
While this finding does not entirely align with our results, many questions about the implications of timescales in these interactions remain open, as the authors themselves note \cite{Zrnic:ANeurIPS:2021}. 
The difference between their and our results may stem from our novel setup, which includes a population of decision-makers, in addition to the population of users.
Hence, here too, our framework may contribute to more clarity about the complex role of response times in strategic classification.

While our model offers valuable insights, we recognize its simplifying assumptions and constraints.

For example, we consider only a limited set of discrete strategies for each kind of player.
For our purpose here, this is sufficient because we also consider only two types of users: Good users --- who would turn out successful in repaying the loan, whether or not they invest in their application --- and Bad users --- who would not turn out successful, unless they invest in honest improvement. 
However, in realistic settings, the user's standing at a given point in time is continuous, being more or less close to a decision boundary.
This would then imply that different users have distinct costs of adaptation \cite{Barsotti:IJCAI:2022}.
Likewise, in principle,  institutions could set their decision threshold at any level.
Moreover, here we define the institution's strategies as the boundary level --- medium or high.
It would be equally interesting to think of strategies that comprise the level of algorithmic transparency.
In such a setup, a more transparent institution conveys the users more accurate information on how the classification algorithm operates.
This allows a more informed user adaptation.
Our algorithmic recourse scenario resonates with this idea.
However, the effects of transparency are still unclear \cite{Ghalme:ICML:2021,Barsotti:IJCAI:2022, Shao:arXiv:2025}, and a full exploration of the strategic aspects of transparency remains to be done. 

Our framework offers a mathematically rigorous way of modelling feedback loops between collectives of users and institutions. Moreover, it naturally allows for testing interventions to mitigate the adverse effects of strategic adaptation. This provides insights on the importance of designing AI models that are not only accurate and robust but also ethically sound.

\section*{Disclaimer}
The views expressed in this paper are
solely those of the authors and do not necessarily represent
the views of their current or previous employers.

\section*{Authors contributions}
Flavia Barsotti: Conceptualization, Methodology, Writing - Original draft preparation, Writing - Reviewing and Editing, Investigation,  Visualization, Supervision, Project administration.
Marta C. Couto: Conceptualization, Methodology, Writing - Original draft preparation, Writing - Reviewing and Editing,  Software, Investigation, Visualization.
Fernando P. Santos: Conceptualization, Methodology, Writing - Original draft preparation, Writing - Reviewing and Editing, Investigation, Visualization, Supervision.

\section*{Conflict of interest}
The authors declare no conflict of interest.

\appendix

\section{Strategy sets simplification}
\label{appendix-strategy-sets}

The full set of strategies available to players would include \textbf{Low}, \textbf{Medium} and \textbf{High} for the institutions, 
and \textbf{Not Adapt}, \textbf{Improve}, and \textbf{Fake} for the populations of both Good and Bad users. In our study, we consider two possible strategies for each of the three populations, i.e. institution, Good users, Bad users. 
This setup is convenient because it leads to a 3-dimensional dynamical system, allowing the full state space to be easily visualized in a single representation.
Here, we argue why this simplification does not compromise the conclusions of the model.

To facilitate the following discussion, we include the complete outcome matrices for the three scenarios considered in the main text.
For the {\textit{baseline scenario}} (section \ref{baseline:imp_class}), the complete matrix would be
\begin{equation}
	\begin{tabular}{  c || c c c | c c c }
		& \multicolumn{3}{c|}{Good} & \multicolumn{3}{c}{Bad} \\
		& $\mathbf{Not\; adapt}$ & $\mathbf{Fake}$ & $\mathbf{Improve}$ & $\mathbf{Not\; adapt}$ & $\mathbf{Fake}$ & $\mathbf{Improve}$ \\
		\hline \hline
		$\mathbf{L}$ & $TP$ & $TP$ & $TP$ & $FP$ & $FP$ & $TP$ \\
		$\mathbf{M}$ & $TP$ & $TP$ & $TP$ & $TN$ & $FP$ & $TP$ \\
		$\mathbf{H}$ & $FN$ & $TP$ & $TP$ & $TN$ & $TN$ & $FN$ \\
	\end{tabular}.
\end{equation}

For the {\textit{manipulation-proof scenario}} (section \ref{manip_class}), the complete matrix would be
\begin{equation}
	\begin{tabular}{  c || c c c | c c c }
		& \multicolumn{3}{c|}{Good} & \multicolumn{3}{c}{Bad} \\
		& $\mathbf{Not\; adapt}$ & $\mathbf{Fake}$ & $\mathbf{Improve}$ & $\mathbf{Not\; adapt}$ & $\mathbf{Fake}$ & $\mathbf{Improve}$ \\
		\hline \hline
		$\mathbf{L}$ & $TP$ & $TP$ & $TP$ & $FP$ & $FP$ & $TP$ \\
		$\mathbf{M}$ & $TP$ & $TP$ & $TP$ & $TN$ & $\textbf{\textit{TN}}$ & $TP$ \\
		$\mathbf{H}$ & $FN$ & $TP$ & $TP$ & $TN$ & $TN$ & $FN$ \\
	\end{tabular}.
\end{equation}

For the {\textit{algorithmic recourse scenario}} (section \ref{algo_rec}), the complete matrix would be
\begin{equation}
	\begin{tabular}{  c || c c c | c c c }
		& \multicolumn{3}{c|}{Good} & \multicolumn{3}{c}{Bad} \\
		& $\mathbf{Not\; adapt}$ & $\mathbf{Fake}$ & $\mathbf{Improve}$ & $\mathbf{Not\; adapt}$ & $\mathbf{Fake}$ & $\mathbf{Improve}$ \\
		\hline \hline
		$\mathbf{L}$ & $TP$ & $TP$ & $TP$ & $FP$ & $FP$ & $TP$ \\
		$\mathbf{M}$ & $TP$ & $TP$ & $TP$ & $TN$ & $FP$ & $TP$ \\
		$\mathbf{H}$ & $FN$ & $TP$ & $TP$ & $TN$ & $TN$ & $\textbf{\textit{TP}}$ \\
	\end{tabular}.
\end{equation}

Let us first look at the strategies available to institutions. 
If we focus only on the matrix concerning Good users, we see that \textbf{Low} strategy is equivalent to \textbf{Medium}, both always accept a Good user, no matter their strategy.
If we focus on the matrix concerning Bad users, the \textbf{Low} strategy is dominated by \textbf{Medium}, in all scenarios, as it renders more false positives.
Therefore, we drop \textbf{Low} to keep our analysis simple, without loss of insight. 
Now we focus on the strategies available to the users.
A Good user, by deciding to \textbf{Not adapt}, keeps their current score and incurs no cost.
In deciding to \textbf{Adapt}, the user increases their score at some cost.
We could interpret this adaptation as truthful (\textbf{Improve}) or not (\textbf{Fake}).
When facing \textbf{High} institutions, they could be tempted to \textbf{Fake} to adapt at a lower cost; still, there would be no damage for the bank in terms of false positives $FP$ as Good users have --- by definition --- the ability to repay the loan. 
In practice, there could even be a risk of being detected, which may prevent Good users from faking.
For our purposes, it only matters whether we observe some costly adaptation (compared to none at no cost), regardless of its type.
For clarity, here we assume this adaptation to be truthful, i.e., equivalent to improvement.
We denote this strategy by \textbf{Adapt} such that there is no ambiguity about the type of player to which we refer.

As for the Bad user, we could allow them to also \textbf{Not Adapt}, that is, not changing their score at no cost.
This strategy would have an advantage compared to \textbf{Fake} or \textbf{Improve} when facing \textbf{High} institutions, for the first two scenarios. 
This is because the user is turned down either way, so the best they could do is to pay no cost.
For the baseline scenario, this would mean that Bad users end up not adapting (when \textbf{High} institutions are present), but this has no consequence on the institutions, nor on the Good users.
As for the manipulation-proof scenario, \textbf{Medium} institutions would still dominate, in which case, to \textbf{Improve} would be the only way to be accepted, and therefore, the preferred action.
Finally, for the algorithmic recourse scenario, \textbf{Not Adapt} would be dominated.

As such, the simplified version of the strategy sets does not restrict the scope of our conclusions.

\section{Linear stability analysis}
\label{appendix-stability}

In this section, we present some details of the local stability analysis, following standard methods (e.g. \cite{Strogatz:book:1994}). 
Our model considers three alternative scenarios described in the main text: baseline scenario (Section \ref{baseline:imp_class}, Appendix \ref{app:base}), manipulation-proof classifier scenario (Section \ref{manip_class}, Appendix \ref{app:manip}) and algorithmic recourse scenario (Section \ref{algo_rec}, Appendix \ref{app:algo}). 
Here, for each of the three scenarios, we (i) define the dynamical system, (ii) find the fixed or rest points of the system, (iii) write the Jacobian matrix, and (iv) compute the respective eigenvalues, from which we derive conditions for (linear) stability.\\

For each scenario, we consider three-dimensional dynamical systems that can be generically written as
\begin{equation}
\begin{aligned}
\dot{x} &= f_1(x, y, z), \\
\dot{y} &= f_2(x, y, z), \\
\dot{z} &= f_3(x, y, z),
\end{aligned}
\label{equation-system}
\end{equation}
\noindent (where in the main text the variable $x$ corresponds to $x_1$, $y$ to $y^G_1$ and $z$ to $y^B_1$).

Let the set of fixed points of the system be denoted by $\{(x^*_i, y^*_i, z^*_i)\}_{i=1}^{N}$.
To find them, we solve
\begin{equation}
f_1(x^*, y^*, z^*) = 0, \quad f_2(x^*, y^*, z^*) = 0, \quad f_3(x^*, y^*, z^*) = 0.
\label{equation-fixedpoints}
\end{equation}

The Jacobian matrix $\mathbf{J}$ of the system is given by

\begin{equation}
\mathbf{J}(x, y, z) = 
\begin{bmatrix}
\displaystyle\frac{\partial f_1}{\partial x} & \displaystyle\frac{\partial f_1}{\partial y} & \displaystyle\frac{\partial f_1}{\partial z} \\
\displaystyle\frac{\partial f_2}{\partial x} & \displaystyle\frac{\partial f_2}{\partial y} & \displaystyle\frac{\partial f_2}{\partial z} \\
\displaystyle\frac{\partial f_3}{\partial x} & \displaystyle\frac{\partial f_3}{\partial y} & \displaystyle\frac{\partial f_3}{\partial z}
\end{bmatrix}.
\label{equation-jacobian}
\end{equation}

For each fixed point $(x^*_i, y^*_i, z^*_i)$, we compute the Jacobian $\mathbf{J}_i = \mathbf{J}(x^*_i, y^*_i, z^*_i)$. 
The local stability of each fixed point $i$ is determined by the eigenvalues $\{\lambda_1, \lambda_2, \lambda_3\}$ of $\mathbf{J}_i$. 
Specifically,

\begin{itemize}
    \item if all eigenvalues have negative real parts, the fixed point is locally asymptotically stable (an attractor),
    \item if at least one eigenvalue has a positive real part, the fixed point is unstable (a repeller or saddle),
    \item if eigenvalues have zero real parts or are purely imaginary, linear stability is inconclusive (center or bifurcation point).
\end{itemize}

We next analyse each case.

\subsection{Baseline: Imperfect classifier}
\label{app:base}

For the baseline scenario (imperfect classifier, Section \ref{baseline:imp_class}), we recall the system in Eq.\eqref{equation-replicator-baseline}, given by:
\begin{equation}\label{}
 \begin{split}
      \dot{x}_1 &= r \; x_1 \big(1-x_1\big)\big(\rho - \rho \, p_G \, (1-y_1^G) - (\lambda+\rho)(1-p_G) \, y_1^B  \big),  \\
\dot{y}_1^G &= y_1^G \big(1-y_1^G\big) \big(c_I - b(1-x_1)\big), \\
\dot{y}_1^B &= y_1^B \big(1-y_1^B\big) \big(c_I - c_F\big).
 \end{split}
\end{equation}

\noindent This system has trivial fixed points at the corners of the state space, that is, $x_1^*=0,1$, $y^{G*}_1=0,1$, and $y^{B*}_1=0,1$.
This results from properties of the replicator dynamics~\cite{Hofbauer:book:1998}, so we observe the same trivial fixed points for the other two scenarios, namely manipulation-proof classifier scenario (Section \ref{manip_class}, Appendix \ref{app:manip}) and algorithmic recourse scenario (Section \ref{algo_rec}, Appendix \ref{app:algo}).
Additionally, there are two non-trivial fixed points.

The Jacobian matrix $\mathbf{J}(x_1, y^G_1, y^B_1)$ is given by
\begin{align*}
\resizebox{\textwidth}{!}{$\displaystyle
\begin{bmatrix}
r \left( 1 - 2 x_1 \right) \left( \rho - \rho \, p_G (1 - y^G_1) - (\lambda + \rho)\,(1 - p_G)\, y^B_1\right)  &  \rho \, r\, p_G\, x_1\,(1 - x_1) & - \left( r \, (\lambda + \rho) \,(1 - p_G)\, x_1 \, (1 - x_1) \right) \\
b \, y^G_1 \, (1 - y^G_1) & -\left((b\, -c_I \, (1 - x_1))\, (1 - 2\, y^G_1)\right) & 0 \\
0 & 0 & (c_I - c_F)\, (1 - 2\, y^B_1)  
\end{bmatrix}$}.
\end{align*}

\noindent We evaluate the Jacobian matrix at each fixed point and calculate its eigenvalues.

From all eigenvalues, we identify only two candidates to be locally stable fixed points: $(0,0,1)$ and $(1,1,1)$.
The eigenvalues for $(0,0,1)$ are
\begin{equation}
\begin{aligned}
\lambda_1 &= c_F-c_I, \\
\lambda_2 &= \frac{1}{2} \left(-b + c_I - \lambda\,r\,(1 - p_G) - \sqrt{(b - c_I - \lambda\,r\,(1 - p_G))^2} \right), \\
\lambda_3 &= \frac{1}{2} \left(-b + c_I - \lambda\,r\,(1 - p_G) + \sqrt{(b - c_I - \lambda\,r\,(1 - p_G))^2} \right).
\end{aligned}
\end{equation}
We can immediately see that $\lambda_1<0$ for our fundamental assumption on the costs of faking and improving, i.e. $c_F<c_I$.
To check whether $\lambda_2$ and $\lambda_3$ are negative, we consider the cases (i) $(b - c_I - \lambda\,r\,(1 - p_G))>0$ and (ii) $(b - c_I - \lambda\,r\,(1 - p_G))<0$, separately.
For case (i), $\lambda_2$ and $\lambda_3$ simplify to
\begin{equation}
\begin{aligned}
\lambda_2 &= c_I - b, \\
\lambda_3 &= - \lambda\,r\,(1 - p_G) .
\end{aligned}
\end{equation}
Therefore, $\lambda_2<0$ and $\lambda_3<0$ for all our assumptions $p_G \in [0,1]$ and $c_I<b$. 

For the case$(b - c_I - \lambda\,r\,(1 - p_G))<0$, the eigenvalues $\lambda_2$ and $\lambda_3$ simplify to
\begin{equation}
\begin{aligned}
\lambda_2 &= - \lambda\,r\,(1 - p_G), \\
\lambda_3 &= c_I - b.
\end{aligned}
\end{equation}
Like in case (i), $\lambda_2<0$ and $\lambda_3<0$ for all our parameter assumptions.

Therefore, $(0,0,1)$ is a linearly stable fixed point.

The eigenvalues for $(1,1,1)$ are
\begin{equation}
\begin{aligned}
\lambda_1 &= c_F-c_I, \\
\lambda_2 &= \frac{1}{2} \left( -c_I + r\,(\lambda\,(1 - p_G)- \rho\,p_G) - \sqrt{(c_I + r\,(\lambda \,(1 - p_G) - \rho \, p_G))^2} \right), \\
\lambda_3 &= \frac{1}{2} \left( -c_I + r\,(\lambda\,(1 - p_G)- \rho\,p_G) + \sqrt{(c_I + r\,(\lambda \,(1 - p_G) - \rho \, p_G))^2} \right).
\end{aligned}
\end{equation}
We can immediately see that $\lambda_1<0$ for our fundamental assumption $c_F<c_I$.
Again, to check whether $\lambda_2$ and $\lambda_3$ are negative, we consider cases (i) $c_I + r\,(\lambda \,(1 - p_G) - \rho \, p_G))>0$ and (ii) $c_I + r\,(\lambda \,(1 - p_G) - \rho \, p_G))<0$, separately.
For case (i), $\lambda_2$ and $\lambda_3$ simplify to
\begin{equation}
\begin{aligned}
\lambda_2 &= -c_I,\\
\lambda_3 &= r\,(\lambda\,(1 - p_G)- \rho\,p_G).
\end{aligned}
\end{equation}
Thus, $\lambda_2<0$.
As for the other eigenvalue, $\lambda_3<0$ only if 
\begin{equation}
    \lambda\,(1 - p_G)- \rho\,p_G < 0 \Rightarrow p_G>\frac{\lambda}{\lambda+\rho},
\end{equation}
which is the condition in Eq.\eqref{equation-fixed-point-condition} of the main text.

In the case $(c_I + r\,(\lambda \,(1 - p_G) - \rho \, p_G))<0$ the eigenvalues for $(1,1,1)$ simplify to
\begin{equation}
\begin{aligned}
\lambda_2 &= r\,(\lambda\,(1 - p_G)- \rho\,p_G),\\
\lambda_3 &= -c_I.
\end{aligned}
\end{equation}
These are the same as above.
We verify that $\lambda_3<0$, and $\lambda_2<0$ only if $p_G>{\lambda}/{(\lambda+\rho)}$.

\subsection{Manipulation-proof classifier}
\label{app:manip}
For the manipulation-proof scenario (Section \ref{manip_class}), we recall the system in Eq.\eqref{equation-replicator-robust}
\begin{equation}
 \begin{split}
      \dot{x}_1 &= r \; \rho \; x_1 \big(1-x_1\big)\big(1 - p_G\,(1-y_1^G) - y_1^B\,(1-p_G)  \big),  \\
\dot{y}_1^G &= y_1^G \big(1-y_1^G\big) \big(c_I - b \, (1-x_1)\big), \\
\dot{y}_1^B &= y_1^B \big(1-y_1^B\big) \big(c_I - c_F - b \, x_1\big).
 \end{split}
\end{equation}

Again, the system has trivial fixed points at all corners of the state space.
Additionally, when $y_1^G=0$ and $y_1^B=1$, then $\dot{x_1}=0$, for any value of $x_1$.
Therefore, all points $(x_1,0,1)$ for $x_1 \in [0,1]$ are fixed points of the system.
They are not necessarily all locally stable.
Since these states do not appear as end states very often when we numerically integrate the system for the parameters of interest, we do not focus our analysis on them.
Besides these points, there are four more non-trivial points.

The Jacobian matrix $\mathbf{J}(x_1, y^G_1, y^B_1)$ is given by
\begin{align*}
\resizebox{\textwidth}{!}{$\displaystyle
\begin{bmatrix}
r \rho (1 - 2 x_1) \left( 1 - p_G(1-y^G_1) - (1-p_G)y^B_1 ) \right) &
 \rho \, r \, p_G \, x_1 \, (1 - x_1) &
 -\rho \, r \, (1 - p_G)\, x_1\, (1 - x_1)\, \\
b\, (1 - y^G_1)\, y^G_1 &
- \left( c_I + b (1 - x_1) \right) (1 - 2 y^G_1) &
0 \\
- b\, (1 - y^B_1)\, y^B_1 &
0 &
\left( c_I - c_F - b\, x_1 \right)(1 - 2 y^B_1)
\end{bmatrix}$}.
\end{align*}

\noindent We evaluate the Jacobian matrix at each fixed point and calculate its eigenvalues.
By evaluating the eigenvalues at parameter values in the range of interest, we identify only one candidate to be locally stable: $(1,1,0)$.
The respective eigenvalues for this point are
\begin{equation}
\begin{aligned}
\lambda_1 &= -c_I, \\
\lambda_2 &= -b - c_F + c_I, \\
\lambda_3 &= -r \, \rho.
\end{aligned}
\end{equation}
These are all negative for our assumptions, since all costs are positive values and $b>c_I$. 
Therefore, $(1,1,0)$ is a linearly stable fixed point.

\subsection{Algorithmic recourse}
\label{app:algo}
Finally, for the algorithmic recourse scenario (Section \ref{algo_rec}), we rewrite the system of equations \eqref{equation-replicator-recourse}
\begin{equation}
 \begin{split}
\dot{x}_1 &= r \; x_1 \big(1-x_1\big) \big(\rho \, p_G\,y_1^G - \lambda \,(1-p_G) \, y_1^B  \big),  \\
\dot{y}_1^G &= y_1^G \big(1-y_1^G\big) \big(c_I - b \, (1-x_1)\big), \\
\dot{y}_1^B &= y_1^B \big(1-y_1^B\big) \big(c_I - c_F - b \, (1-x_1)\big).
 \end{split}
\end{equation}

\noindent When $y_1^G=y_1^B=0$, then $\dot{x_1}=0$, for any value of $x_1$.
Therefore, all points $(x_1,0,0)$ for $x_1 \in [0,1]$ are fixed points of the system.
They are not necessarily all locally stable.
Besides these, there are two more non-trivial points.

The Jacobian matrix $\mathbf{J}(x_1, y^G_1, y^B_1)$ is then given by
\begin{align*}
\resizebox{\textwidth}{!}{
$\displaystyle\begin{bmatrix}
r \left( 1 - 2 x_1 \right) \left( \rho \, p_G \, y^G_1 - \lambda \, (1 - p_G) \, y^B_1\right)  &  \rho \, r\, p_G\, x_1\,(1 - x_1) & - r \, \lambda \,(1 - p_G)\, x_1 \, (1 - x_1) \\
b \, y^G_1 \, (1 - y^G_1) & (c_I\, -b \, (1 - x_1))\, (1 - 2\, y^G_1) & 0 \\
b \, y^B_1 \, (1 - y^B_1) & 0 & (c_I - c_F - b(1-x_1))\, (1 - 2\, y^B_1) & 
\end{bmatrix}$}.
\end{align*}

The fixed point $(1,1,1)$ has eigenvalues
\begin{equation}
\begin{aligned}
\lambda_1 &= -c_I, \\
\lambda_2 &= c_F - c_I, \\
\lambda_3 &= r\,(\lambda\,(1 - p_G)- \rho\,p_G).
\end{aligned}
\end{equation}
Thus, like for the baseline scenario, $(1,1,1)$ is stable if $p_G>\lambda/(\lambda+\rho)$.

We also notice that the fixed point $(\frac{b+c_F-c_I}{b},1,\frac{\rho \, p_G}{\lambda \, (1-p_G)})$ has eigenvalues $\lambda_1 = -c_F$, and $\lambda_i$ with $i=2,3$ such that 
\begin{equation}
\lambda_i^2-\frac{r\, \rho\, p_G\, (c_F - c_I)(b + c_F - c_I)\, \left((1 - p_G)\lambda - p_G\, \rho\right)}{b (1 - p_G)\lambda}=0.
\end{equation}
For $p_G<\lambda/(\lambda+\rho)$, the solutions $\lambda_i$ of this equation are purely imaginary.
Therefore, the fixed point $(\frac{b+c_F-c_I}{b},1,\frac{\rho \, p_G}{\lambda \, (1-p_G)})$ corresponds to cyclic dynamics.

\section{Additional results on cyclic dynamics}
\label{appendix-cycles}
\begin{figure}[h!]
    \centering
    \includegraphics[width=0.9\linewidth]{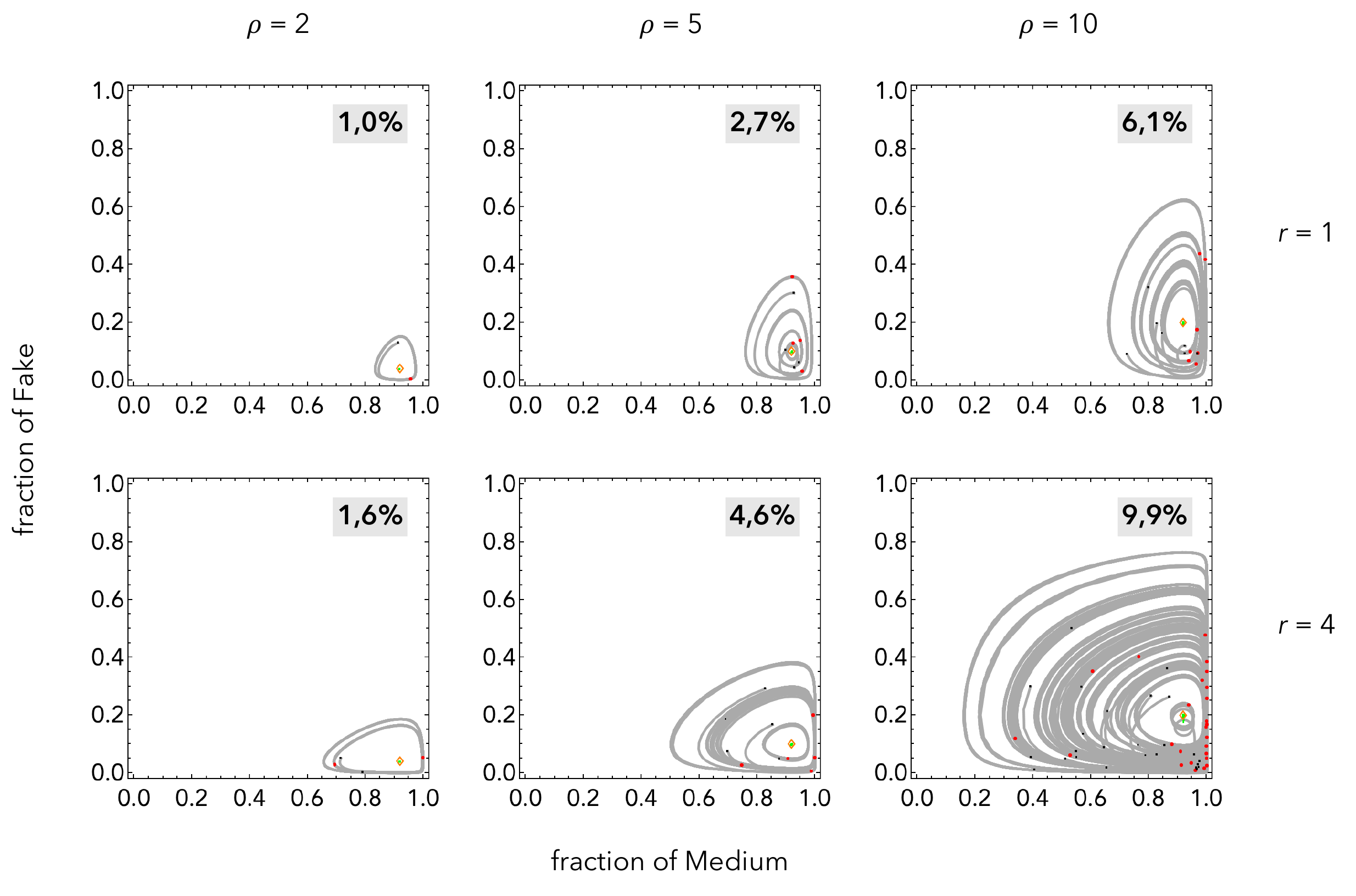}
    \caption{\textbf{Cycling dynamics in the algorithmic recourse scenario.} 
    Since the cycles occur only among the institutions and the Bad users (the Good users ending up fully in \textbf{Not adapt}), here, we plot a projection of the cycling trajectories onto the respective 2-dimensional state space.
    The x-axis corresponds to the fraction of strategy \textbf{Medium} and the y-axis to the fraction of strategy \textbf{Fake}.
    We numerically solve the differential equation given by Eq.\eqref{equation-replicator-recourse} starting from $200$ random initial conditions over the entire state space.
    We plot only the trajectories that correspond to cycles.
    Therefore, the number of trajectories shown correlates with the basins of attraction.
    For a more accurate measure, we also present the percentage of initial conditions that lead to cycles (gray squares) out of $8000$ (similarly to how we compute basins of attraction in the main text).
    The initial ($t=0$) and end points ($t=50$) of the trajectories are shown in black and red, respectively.
    The average value over time for each trajectory (numerically computed) is shown in green circles.
    For comparison, we indicate the location of the fixed point $(\frac{b+c_F-c_I}{b},1,\frac{\rho \, p_G}{\lambda \, (1-p_G)})$ with an orange diamond, verifying that they closely match.
    We present the results for $\rho=2, 5, \mbox{and } 10$ and $r=1,\mbox{and } 4$. The other parameters are fixed: $\lambda=50$, $b=50$, $c_F=1$, $c_I=5$, $p_G=0.5$. 
    \label{fig-si1}}
\end{figure}

Here, we further explore the cyclic regime of the algorithmic recourse scenario (Section \ref{algo_rec}).
Namely, we show how the shapes and basins of attraction of the cycles vary with parameters $\rho$, the benefit of a true positive for the institution, and $r$, the adaptation rate of the institutions relative to users (Fig.~\ref{fig-si1}).
Since cycles occur only among the institutions and the Bad users (the Good users ending up fully in {\textbf{Not adapt}}), here, we plot a projection of the periodic trajectories onto the respective 2-dimensional state space.
We verify that the larger $\rho$ and $r$, the more initial conditions lead to cycles (denoted by the percentages in gray squares).
At the same time, we note that the higher $\rho$, the higher is the fraction of \textbf{Fake} on average (thus, the lower performance).
Therefore, $\rho$ has a delicate effect on the dynamics, promoting cycles while decreasing performance.
On the other hand, increasing $r$ also increases the basin of attraction of cyclic orbits, while keeping the average behavior over the cycles constant (Fig.~\ref{fig-si1} top to bottom panels). 
We can also confirm this by the non-dependence of $r$ on the fixed point $(\frac{b+c_F-c_I}{b},1,\frac{\rho \, p_G}{\lambda \, (1-p_G)})$.

\newpage

\bibliographystyle{ACM-Reference-Format}

\end{document}